\documentclass[%
 reprint,
 amsmath,amssymb,
 aps,
]{revtex4-2}
\usepackage{placeins}
\usepackage{graphicx}
\usepackage{dcolumn}
\usepackage{xcolor}
\usepackage{bm}

\newcommand{\xv}{{\bf x}}

\begin{document}

\preprint{APS/123-QED}

\title{Geometrically Frustrated Assembly at Finite Temperature: Phase Transitions from Self-Limiting to Bulk States}

\author{Nicholas W. Hackney}
 \email{nwhackn@sandia.gov}
\affiliation{%
 Sandia National Laboratories, Albuquerque, New Mexico 87185, USA\\
}%

\author{Gregory Grason}
\affiliation{
Polymer Science and Engineering Department, University of Massachusetts Amherst\\
}%

\date{\today}

\begin{abstract}
Geometric frustration is recognized to generate complex morphologies in self-assembling particulate and molecular systems. In bulk states, frustrated drives structured arrays of topological defects.   In the dilute limit, these systems have been shown to form a novel state of self-limiting assembly, in which the equilibrium size of multi-particle domains are finite and well-defined. In this article, we employ Monte Carlo simulations of a recently developed 2D lattice model of geometrically frustrated assembly~\cite{HackneyPhysRevX.13.041010} to study the phase transitions between the self-limiting and defect bulk phase driven by two distinct mechanisms: (i) increasing concentration and (ii) decreasing temperature or frustration.  The first transition is mediated by a concentration-driven percolation transition of self-limiting, worm-like domains into an intermediate heterogeneous network mesophase, which gradually fills in at high concentration to form a quasi-uniform defect bulk state. We find that the percolation threshold is weakly dependent on frustration and shifts to higher concentration as frustration is increased, but depends strongly on the ratio of cohesion to elastic stiffness in the model. The second transition takes place between self-limiting assembly at high-temperature/frustration and phase separation into a condensed bulk state at low temperature/frustration.  We consider the competing influences that translational and conformational entropy have on the critical temperature/frustration and show that the self-limiting phase is stabilized at higher frustrations and temperatures than previously expected. Taken together, this understanding of the transition pathways from self-limiting to bulk defect phases of frustrated assembly allows us to map the phase behavior of this 2D minimal model over the full range of concentration.

\end{abstract}

\maketitle


\section{Introduction}\label{section: intro}

Geometric frustration refers to the scenario in which global geometrical constraints on the degrees of freedom of an interacting system obstruct the uniform states of locally preferred order across the entire system~\cite{sadoc}. This idea has been used to rationalize the behavior of a wide array of systems ranging from low-temperature magnetism~\cite{vannimenus1977theory,WannierPhysRev.79.357} to superconductors in an external field~\cite{franz1995vortex}. Further, geometric frustration has been recognized to play an important role in shaping the behavior of many different systems of self-assembling particles in soft matter~\cite{grason2016perspective, witten, meiri2021cumulative, HaganGrason}. Here, deformation of inter-particle arrangement away from the locally preferred state can result in intra-assembly strain gradients that propagate to long range. Examples of such geometrically frustrated assembly include crystallization on curved surfaces~\cite{bausch2003grain,schneider2005shapes,irvine2010pleats,meng}; non-tiling polygonal~\cite{witten,roy2023collective,spivack2022stress} and polyhedral particles~\cite{MaoPhysRevLett.131.258201,serafin}; twisted bundles of self-assembling protein filaments~\cite{aggeli2001hierarchical,hall2016morphology}; hyperbolic membranes~\cite{GhafBruin,armon2014shape,Sharon, tyukodi2022thermodynamic,Hall2023}; liquid crystals on curved surfaces~\cite{vitelli2006nematic,Nieves2007,lopez2011frustrated,carenza2022cholesteric}; two-dimensional assemblies of bent-core nematic liquid crystals~\cite{NivEfrati,fernandez2021hierarchical}; one-dimensional assemblies of curved colloidal~\cite{tanjeem2022focusing, sullivan2024self} and incommensurate `polybrick' particles~\cite{Berengut2020,wang2024thermal}; as well as models of viral capsid assembly~\cite{Mendoza,li2018large}.

Broadly speaking, investigations into the behavior of these systems have focused on the case of either macroscopic bulk or dilute self-assembly. In the first case, the assembling subunits extend to completely fill the available space. Here, the lack of both translational degrees of freedom and free boundaries limit the systems ability to alleviate the accumulation of intra-assembly strain gradients, leading to the proliferation of topological defects throughout the ground state~\cite{sadoc,kleman1989curved, tarjus2005frustration}. Conversely, in systems of dilute self-assembly, the added subunit mobility allows for relaxation of the effects of frustration via variation of the size and shape of the individual multi-unit domains. These extra degrees of freedom have been argued to give rise to unique scale-dependent behavior known as self-limiting assembly~\cite{HaganGrason} resulting from the thermodynamic balance between cohesive loss at the boundary and elastic cost of accumulating stress due to frustration.  Conceptually, the putative self-limiting state is of particular interest as it implies that frustrated assemblies are able to ``measure'' and limit their equilibrium dimensions at size scales much larger than the individual subunits or their interaction range.  This property is also of potential value for creating synthetic self-assembling materials that can target well-defined mesostructured dimensions that are tunable through programmable misfit of the constituent subunits themselves.

 While early theories of self-limitation (e.g. \cite{schneider2005shapes, Turner2003}) have been derived from ground-state thermodynamics of continuum elastic models of frustrated assembly, the thermodynamic stability of the self-limiting state at finite temperature has only more recently been studied~\cite{HackneyPhysRevX.13.041010, wang2024thermal}. Here it was shown that, for fixed and sufficiently dilute concentration, the self-limiting state occurs at intermediate frustration strength. In addition, this state was shown to be flanked by the dispersed state at high frustration and a defect-riddled bulk state at low, but non-zero frustration. In this article, we focus on the possible thermodynamic pathways between the self-limiting and bulk states more broadly, and in the context of a recently developed 2D lattice model of geometrically frustrated assembly~\cite{HackneyPhysRevX.13.041010}.  The model is an extension of the uniformly frustrated 2D XY model~\cite{tarjus2005frustration, Esterlis} that incorporates translational degrees of freedom at states of fixed total subunit concentration, which is known to encode the minimal elements of frustration of orientational 2D order on surfaces with fixed and non-zero Gaussian curvature~\cite{nelson1987fluctuations, bowick}.  In particular, it was shown that self-limiting states of this generalized lattice model take the form of finite-width, worm-like domains, while the bulk condensed state takes the form of a sponge-like domain where the ``holes'' correspond to topologically charged vortices that screen frustration.  

While Monte Carlo simulations of the frustrated lattice model have demonstrated the existence of the self-limiting state at finite-temperature and characterized a transition to a bulk-condensed state below a critical frustration strength, two key aspects of the self-limiting to bulk transition remain unclear.  First, the existence of a self-limiting to bulk transition with decreasing frustration can be argued purely on the ground state energetics of these states and their distinct frustration dependence.  This energetic scaling argument suggests a critical frustration value for the transition proportional to the square of the cohesion to (spin) stiffness ratio, notably {\it independent} of temperature.  While the critical frustration observed from simulations was shown to increase with cohesion to stiffness, it was also found to be strongly temperature dependent, which implies that entropic factors in the free energy difference between bulk and self-limiting states are non-negligible.  At present, it is not clear what is the most significant source of entropy difference between these two states (e.g. translational or conformational fluctuations), how it compares to the relative cohesive and elastic energetic difference, and if and under what conditions this transition ever approaches the purely energetic description.  A second, and far more open question, is how the phase behavior of the frustrated lattice model evolves at fixed, intermediate frustration with {\it increasing concentration}. While at dilute conditions, the self-limiting state is formed above a (pseudo-)critical aggregation concentration (CAC), it is intuitive to expect a bulk equilibrium state in the limit of large concentration, where the translational degrees of freedom are severely limited and excluded volume effects become strong.  It is natural to expect that this high-concentration bulk state will be a space-filling version of the defect-riddled, spongey condensate that forms below the critical value of frustration in dilute conditions (i.e. a semi-regular array of vortices with voided ``cores'').  However, it is so far unclear how the system proceeds thermodynamically from a state of dispersed, finite-width and variable length aggregates to a solid-like and quasi-uniform array of defect voids.  The existence of one or more possible mesophases (e.g. liquid crystalline order) intermediate to self-limiting and bulk states, and the corresponding nature of the thermodynamic transitions between such states, is not known.

\begin{figure*}[ht]%
\includegraphics[width=\textwidth]{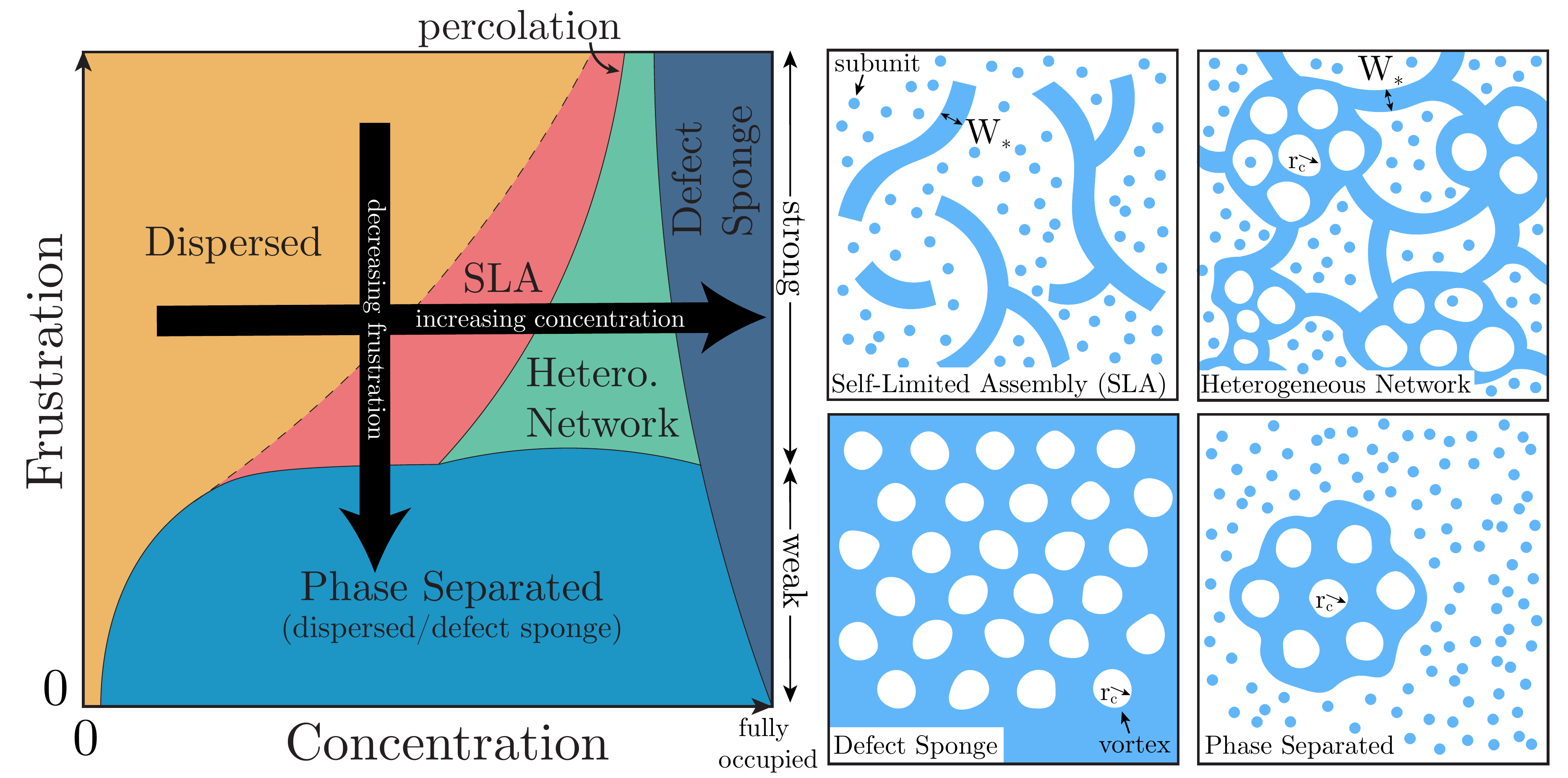}
\caption{Sketch of the important features in the frustration-concentration plane of the phase diagram of geometrically frustrated assembly. The concentration axis ranges from $\Phi=0.0$ (empty lattice) to $\Phi=1.0$ (full lattice). The frustration axis is depicted over the entire range of physically meaningful frustrations values, with $f=0$ corresponding to no frustration and $f=1/2$ corresponding to full frustration. Illustrations are provided for each of the important states. }\label{fig: cartoon phase diagram}
\end{figure*}


In this article, we address these open aspects of the self-limiting to bulk transition in a minimal 2D lattice model of geometrically frustrated assembly~\cite{HackneyPhysRevX.13.041010} using a combination of Monte Carlo simulation and  mean-field, continuum theory modeling of its assembly thermodynamics.  The scope and key findings of this study are illustrated schematically in the phase diagram shown in Figure \ref{fig: cartoon phase diagram} along with corresponding cartoons of the different competing phases.  First, for the concentration driven transition at fixed intermediate frustration (horizontal arrow in Fig.~\ref{fig: cartoon phase diagram}), we show the existence of a phase transition between self-limiting assembly and an intermediate mesophase characterized by percolation of finite-width domains into a heterogeneous ``gel-like" network. We show that this percolation transition is weakly dependent on frustration, but strongly dependent on the ratio of cohesion to stiffness, behavior which we compare to an effective bond percolation of self-limiting domains.   Upon further increase in concentration, the aggregated and disordered network continues to fill in and continuously evolves to a ``quasi-uniform" bulk sponge state that consists of a regular array of similarly-sized, evenly spaced holes surrounding topologically charged vortices.  In the limit of full surface coverage, these holes are then uniformly filled in to reach the fully occupied bulk, Abrikosov defect ground state of the uniformly frustrated XY model. We then determine the optimal density of the defect sponge phase, which characterizes the transition between disordered and ordered bulk sponge states, and show that it decreases with frustration and the ratio of inter-particle cohesion to interaction stiffness. 
Next, we revisit the frustration-driven transition between self-limiting and bulk states at fixed dilute concentrations (vertical arrow in Fig.~\ref{fig: cartoon phase diagram}). As indicated in the schematic phase diagram, this transition takes place between a poly-disperse collection of finite sized, worm-like aggregates and a phase separated defect bulk. We analyze distinct microstructural contributions to the relative entropy of self-limiting aggregates and bulk condensates, and propose a mean-field argument to capture the expected leading order contributions from the relatively higher entropy of self-limiting states, and their impact on the relative free energy differences between these states.  We show that this model captures the qualitative effect observed in simulation, with the critical frustration {\it decreasing} as temperature is increased.  We further predict the range of thermodynamic conditions at which the relative thermodynamics of self-limiting and bulk defect states is determined by purely energetic considerations of intra-domain elasticity and cohesion and discuss the implications of this regime for accessible conditions for the state of self-limiting assembly. 

The rest of this article is organized as follows. In Section \ref{section: minimal model}, we provide a brief review of the 2D lattice model of frustrated assembly, including the microscopic degrees of freedom and key parameters. In Section \ref{section: continuum model}, we discuss results from the continuum limit of the two key competing morphologies, and describe the connection with microscopic parameters of the model with key length scales that characterize the mesoscopic effects of frustration. 
In Section \ref{section: percolation}, we investigate the existence of a frustration dependent percolation transition from self-limiting assembly to a heterogeneous network sponge and examine how this intermediate state approaches the homogeneous sponge at near full concentration. We also compare the internal structure of this sequence of states to that of the phase-separated defect bulk found at weak frustration. In Section \ref{section: condensation}, we revisit the phase separation that occurs at dilute concentration (i.e. below the percolation threshold) and weak frustration; analyzing several different contributions to the total aggregation entropy, including the translational and configurational entropy of worm-like domains.  We then describe our mean-field model for the leading temperature-dependence of critical frustration.  Finally, in Section \ref{section: conclusion}, we summarize our key results and discuss their implications for the global phase behavior of the frustrated lattice model for the case of fixed, uniform frustration studied here as well as for connections to experimental systems.

\begin{figure*}[ht!]%
\includegraphics[width=\textwidth]{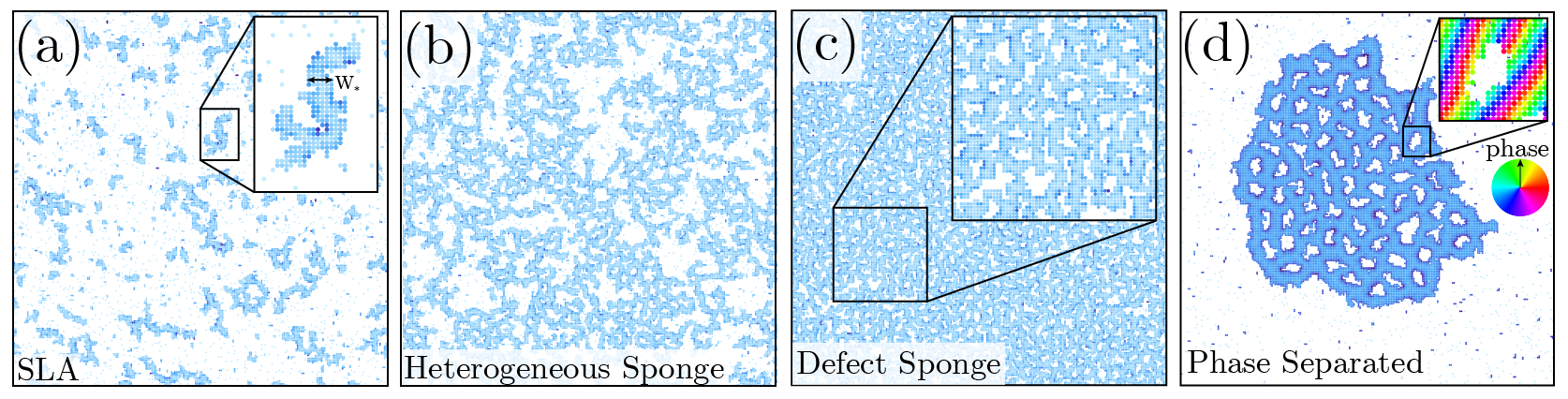}
\caption{Simulation snapshots illustrating each of the important phases of geometrically frustrated assembly. The SLA, heterogeneous sponge and bulk defect sponge simulations were run with the following parameters: $f=0.016$, $\Sigma/J=0.09$, $\beta J=40.0$, $L=250$ and $\Phi=0.20,0.50,0.85$, corresponding to  (a-c) respectively. The phase separated simulations (d) were run with the parameters: $f=0.004$, $\Sigma/J=0.09$, $\beta J=40.0$, $L=500$ and $\Phi=0.10$, with the inset highlight the phase vortex that surrounds holes (defect cores) in the bulk state. Panels show only a $250\times 250$ window of the full simulation box in order to match the scale of the other images.}\label{fig: example simulations}
\end{figure*}
\section{Lattice Model of Geometrically Frustrated Assembly}\label{section: minimal model}
The model of geometrically frustrated assembly that we will be using throughout this article describes a collection of $N$ subunits occupying a fixed fraction, $\Phi=N/L^2$, of sites on an $L\times L$ square lattice with bond spacing $a$. In addition to translational degrees of freedom, each subunit possesses a continuous orientational degree of freedom, $\theta_i\in[0,2\pi]$, i.e. an XY-spin or ``phase" variable. The energetics of this model can be described by the lattice Hamiltonian~\cite{HackneyPhysRevX.13.041010}:
\begin{equation}\label{eq: lattice hamiltonian}
H=-J\sum_{\langle ij\rangle}\cos(\Delta\theta_{ij}-A_{ij})\eta_i\eta_j-K\sum_{\langle ij\rangle}\eta_i\eta_j
\end{equation}
where $\langle ij\rangle$ denotes nearest neighbor lattice sites, $\Delta\theta_{ij}=\theta_i-\theta_j$ defines the nearest neighbor phase difference and $\eta_i=0,1$ denotes the occupancy of each lattice site. The first term describes an $XY$-like interaction based off the relative phase difference between subunits and is parameterized by the spin stiffness $J$. The second term describes an Ising-like nearest neighbor interaction parameterized by the cohesive energy $K$. Lastly, the $A_{ij}$ term represents a gauge field along the lattice links that defines the locally preferred nearest neighbor rotation, and crucially introduces the geometric frustration of an aggregated cluster of spins. Following the standard approach for the uniformly frustrated XY model~\cite{teitelprb1983, tarjus2005frustration}, we define the gauge field subject to the constraint
\begin{equation}
\sum_{\circlearrowright}A_{ij}\equiv 2\pi f.
\end{equation}
Here, $f$ is a measure of the {\it microscopic frustration strength} and the sum is taken over clockwise oriented minimal lattice plaquettes. This constraint implies that, when $f \neq n$ where $n$ is an integer, propagation of the preferred nearest neighbor rotation, $\Delta\theta_{ij}=A_{ij}$, around the plaquette leads to a net angular mismatch of $2\pi f$, and hence the locally preferred state is incommensurate with the existence of closed loops and the system is said to be geometrically frustrated. The parameter $f\in[0,1/2]$ defines the degree to which the occupied square lattice is frustrated, with $f=0$ corresponding to the unfrustrated case and $f=1/2$ corresponding to the case of maximal frustration~\cite{teitelprb1983}.   In this article, we focus on the case of fixed and uniform frustration, where the discrete gauge field can be defined by the integral $A_{ij}=\int_{\mathbf{x}_j}^{\mathbf{x}_i}\rm d\mathbf{x}\cdot\mathbf{A}(\mathbf{x})$ where $\mathbf{A}(\mathbf{x})$ is a two-dimensional vector field defined to have constant curl 
\begin{equation}
\nabla_{\perp}\times\mathbf{A}(\mathbf{x})=2\pi \varphi,
\end{equation}
and $ \varphi\equiv f/a^2$ defines a {\it frustration density}.  Notably, as one considers occupied domains of increasingly larger size (i.e. upon cohesive assembly into compact clusters), the number of incompatible loops grows as well as the magnitude of phase mismatch around the larger loops (i.e. per Stokes law), implying that the energetic cost of the ground state spin configuration will also grow due to an increasing number and magnitude of local spin misalignment.  This {\it accumulation} of energetic cost with increasing domain size has been identified as a hallmark behavior of geometrically frustrated assembly~\cite{meiri2021cumulative} and has been shown to give rise to both the self-limiting and defect-riddled bulk states of assembly.  In terms of specific physical systems~\cite{HackneyPhysRevX.13.041010}, this minimal model most closely describes the frustration of in-plane orientational order (i.e. liquid crystalline) on 2D surfaces of fixed Gaussian curvature~\cite{nelson1987fluctuations,vitelli2006nematic}, proportional to $\varphi$.  

The behavior of this model is determined by two microscopic energy scales.  The first is defined by the net cohesion of an ideally-aligned bond between subunits
\begin{equation}
\Sigma = J +K ,
\end{equation}
while the second is simply $J$ itself, which plays an additional role of spin stiffness, penalizing orientational strain gradients.   The behavior of this lattice model has been studied by MC simulation in the dilute limit ($\Phi\lesssim 0.2$) and for the case of weak cohesion to stiffness ratios, $\Sigma/J \ll 1$~\cite{HackneyPhysRevX.13.041010}. In this regime, there exists an apparent binodal in the frustration-concentration plane below which the system phase separates into a topologically defective condensate surrounded by a dispersed monomer gas. Above the binodal (characterized by a critical frustration value), there is a pseudo-critical aggregation transition from a dispersed gas to a state of self-limiting assembly characterized by distribution of uniform finite width, variable length aggregates in coexistence with dispersed monomers. These aggregates exhibit highly anisotropic ribbon-like morphologies with uniform width controlled by the ratio of $f$ and $\Sigma/J$ and is independent of any extensive properties of the system. For a given value of $\Sigma/J$, the critical frustration along the binodal is apparently constant and independent of concentration. As mentioned in the introduction, this motivates the terminology {\it strong} versus {\it weak} frustration as a means of classifying whether a given set of conditions respectively resides above or below the phase separation binodal.

Conversely, in the fully occupied limit ($\Phi=1$), eq. (\ref{eq: lattice hamiltonian}) described uniformly frustrated 2D XY model, which has been used as a statistical framework for studying field-frustrated superconducting arrays~\cite{teitelprb1983,teitelprl1983,franz1995vortex} and various frustration based glass models~\cite{Esterlis,tarjus2005frustration}. In the fully occupied regime, sufficiently below the Kosterlitz-Thouless temperature of the unfrustrated XY model (i.e. $\beta J \gg 1$), frustration leads to ground states composed of a uniform density of like-signed vortices arranged in a triangular Abrikosov lattice with inter-defect spacing~\cite{alba2008uniformly} 
\begin{equation}\ell_{\rm v} \approx \varphi^{-1/2}.
\end{equation}
In this this article, we apply the Markov chain Monte Carlo algorithm detailed in Ref. \cite{HackneyPhysRevX.13.041010} to study the phase behavior of frustrated assembly across the full range of concentration. Simulation snapshots of each of the unique states of assembly -- self-limiting aggregates, heterogeneous network, quasi-uniform defect sponge and phase separated bulk (sponge) -- enumerated in the schematic phase diagram are provided in Fig. \ref{fig: example simulations}.  Supplementary videos 1-4 show animations of simulated sequences of each of these states.

\section{Continuum Models of Competing Ground State Morphologies}\label{section: continuum model}

Here we describe continuum theory predictions for two characteristic morphologies we observed, shown schematically in Fig. \ref{fig: hole schematic}ab.  This analysis is based on the assumption that, for conditions relevant to self-limiting assembly, thermal fluctuations of spin degrees of freedom are relatively weak.  As we see below, this is justified by the fact that self-limiting domain sizes larger than a single subunit require $\Sigma/J \ll 1$ while equilibrium assembly necessarily requires sufficiently cohesive interactions $\Sigma \gtrsim k_B T$. These conditions imply $J/k_B T \gg 1$ and that spin degrees of freedom of bound clusters can be modeled in terms of their ground state configurations. 

When the phase strain, $\Delta\theta_{ij}-A_{ij}$, is small, we can take the continuum limit of equation (\ref{eq: lattice hamiltonian}) and write a function for the energy of a domain, $\mathcal{D}$, of assembled subunits as:
\begin{equation}\label{eq: continuum Hamiltonian}
\mathcal{H}[\mathcal{D}]=\frac{J}{2}\int_{\mathcal{D}}\rm d^2\mathbf{x}\vert \nabla\theta-\mathbf{A}(\mathbf{x})\vert^2+\sigma\mathcal{P}[\mathcal{D}]-\epsilon_{\rm bulk}\mathcal{A}[\mathcal{D}]
\end{equation}
where the first term describes the elastic energy of deforming the phase field away from its locally preferred state and the last two terms describe the cohesive energy of a domain with perimeter $\mathcal{P}[\mathcal{D}]$ and area $\mathcal{A}[\mathcal{D}]$. Here,  we define the 
\begin{equation}
    \sigma = \frac{\Sigma}{2 a}
\end{equation}
as the line energy of the domain boundary (i.e. cohesive cost per unit length) and $\epsilon_{\rm bulk}=2\Sigma/a^2$ as the cohesive bulk energy density.  Assuming a given fixed domain, $\mathcal{D}$, the Euler-Lagrange equation of for spin degrees of freedom is simply
\begin{equation}\label{eq: euler-lagrange}
\nabla_{\perp}^2\theta(x,y)=\nabla_{\perp}\cdot\mathbf{A}
\end{equation}
with the free boundary conditions
\begin{equation}
\hat{n}\cdot\nabla_{\perp}\theta\vert_{\partial D}=\hat{n}\cdot\mathbf{A} .
\end{equation}
For simplicity we consider a divergence-free gauge
\begin{equation}
\mathbf{A}=\pi\varphi[y\hat{x}-x\hat{y}]
\end{equation}
for which the right hand side of eq. (\ref{eq: euler-lagrange}) vanishes, so that the ground states, $\theta_*$, are simply harmonic functions satisfying the appropriate phase winding condition at the free boundaries. 

For the purposes of understanding the states of optimal assembly, it is necessary to determine the size, shape and topology of domains that optimize the assembly {\it energy density}~\cite{HaganGrason}, 
\begin{equation}
\epsilon[\mathcal{D}]=\frac{J}{2 \mathcal{A}[\mathcal{D}]}\int_{\mathcal{D}}\rm d^2\mathbf{x}\vert \nabla\theta_*-\mathbf{A}(\mathbf{x})\vert^2+\sigma \frac{\mathcal{P}[\mathcal{D}]}{ \mathcal{A}[\mathcal{D}]} - \epsilon_{\rm bulk} .
\end{equation}
Optimal states of assembly of the continuum model derive from optimization of this energy density with respect to domains, and can therefore only depend on two length scales:  the length scale set by frustration density, i.e. $\ell_{\rm v} = \varphi^{-1/2}$ and the second determined by the ratio,
\begin{equation}
\ell_{\rm c} = (\sigma/J)^{-1}
\end{equation}
which constitutes a {\it coheso-elastic} length scale for the assembly.  We now summarize how these length scales determine the structure and energetics of finite-width ribbons and sponge-like, bulk vortex arrays.

\subsection{Finite-width ribbon domains \label{sec: ribbons}}

In reference \cite{HackneyPhysRevX.13.041010}, we computed the assembly free energy (per subunit) landscape from the continuum theory of this lattice frustration model for rectangular, defect free domains of arbitrary cross-sectional dimensions.  This theory showed that optimal ground-state morphologies break symmetry in highly-anisotropic ribbon-like domains, of finite, self-limiting width and arbitrarily unlimited length.  This effect of frustration, which is sometimes referred to as ``filamentation''~\cite{meiri2021cumulative}, favors elastic misfit gradients distributed across the narrow dimension of the domain and uniform elastic energy along the length of the aggregate and is widely observed in 2D frustrated assemblies~\cite{schneider2005shapes, meng, armon2014shape, grason2016perspective, spivack2022stress, serafin, Leroy2023}. Hence, we only review the energetic selection of the finite width, ribbon-like domains.

\begin{figure}[ht!]
\centering
\includegraphics[width=0.45\textwidth]{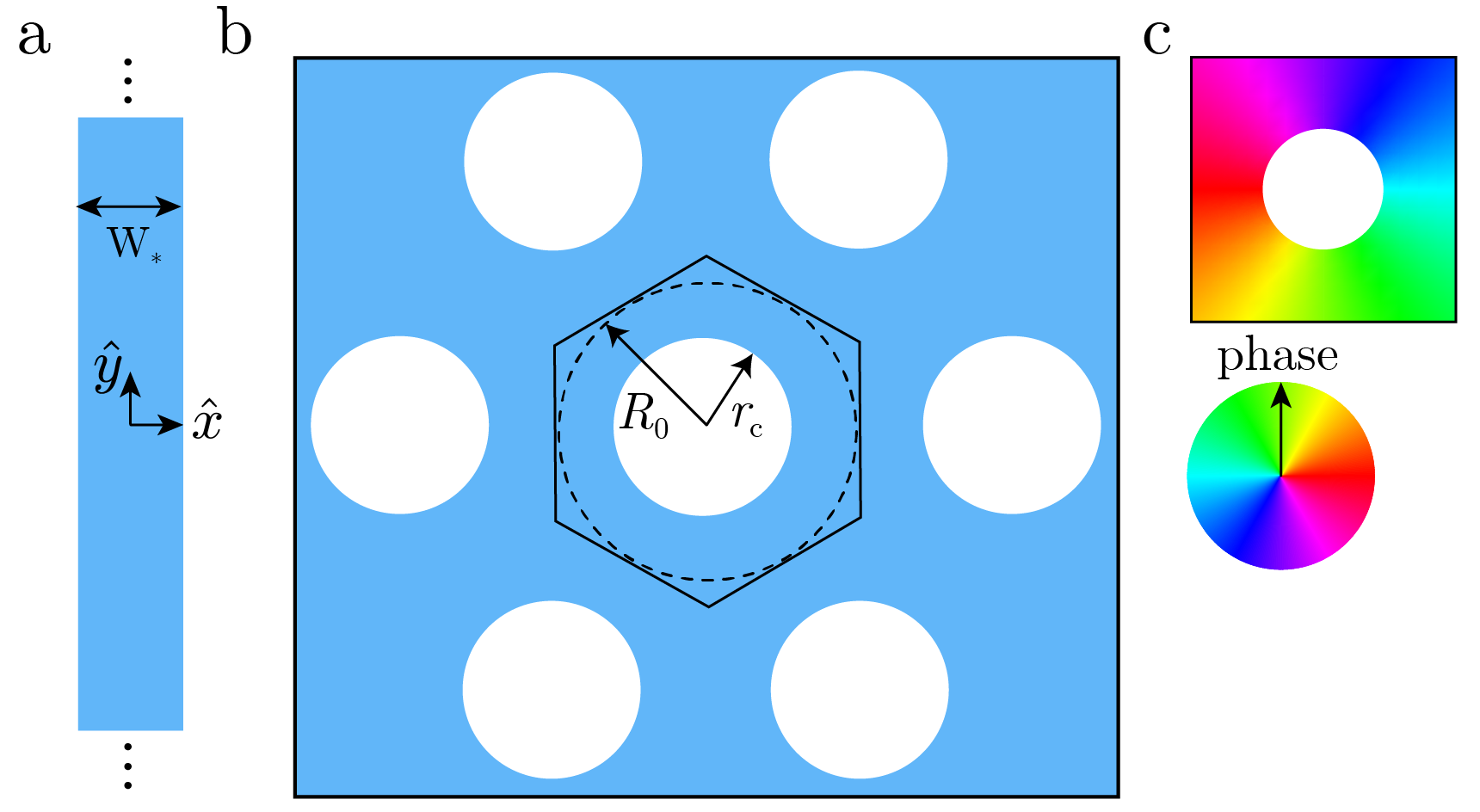}
\caption{Schematic illustration of our model for the (\textbf{a}) self-limiting assembly phase, where we treat the aggregates as infinite strips with uniform width, $W_*$ and (\textbf{b}) the bulk defect sponge phase, which we expect to be a hexagonal lattice of uniformly sized defect holes with radius $r_{\rm c}$. The dashed line depicts the approximate unit cell around each defect hole, which we model as a circle with radius $R_0$. (\textbf{c}) Phase field of the aggregate interior around an individual defect hole is shown to illustrate the winding of the internal rotational degree of freedom. }\label{fig: hole schematic}
\end{figure}

We consider a domain extending along the $y$-axis, and finite width, $W$, in the $x$-direction from $x \in [-W/2,+W/2]$, as shown schematically in Fig. \ref{fig: hole schematic}a.  In this geometry, the boundary conditions require that $\partial_x \theta = \pi \varphi y$ at $x = \pm W/2$.  Additional, uniformity in energy density along the ribbons requires $\partial_y(\nabla \theta -\mathbf{A}) = 0$.  These conditions are satisfied by the harmonic function for ground state phase configuration
\begin{equation}
\theta_*^{\rm (ribbon)} ({\bf x}) =  \pi \varphi x y .
\end{equation}
In finite width ribbons, ground-state phase-strain, $\nabla \theta -\mathbf{A} =  2 \pi \varphi x \hat{y}$, is in the longitudinal direction with transverse gradients in magnitude, and the elastic energy density of ribbons grows quadratically with width, proportional to $J \varphi^2 W^2$.  The full calculation gives an energy density (relative to cohesive energy $-\epsilon_{\rm bulk}$):
\begin{equation}\label{eq: sla energy, dimensions}
\Delta\epsilon(W)\equiv \epsilon(W) + \epsilon_{\rm bulk} = \frac{J\pi^2\varphi^2}{6}W^2+\frac{2\sigma}{W}.
\end{equation}
Minimizing with respect to width gives
\begin{equation}
W_* =\bigg(\frac{6}{\pi^2}\bigg)^{1/3} \ell_{\rm d}
\end{equation}
where we have defined the {\it domain scale}
\begin{equation}\label{eq: ld}
\ell_{\rm d}=\bigg(\frac{\sigma}{J\varphi^2}\bigg)^{1/3} = \ell^{4/3}_{\rm v} \ell^{-1/3}_{\rm c} 
\end{equation}
which is itself a combination of length scales set by the frustration density ($\ell_{\rm v}$) and coheso-elastic interplay ($\ell_{\rm c}$). Inserting the selected width into $\Delta\epsilon(W)$, we obtain the ground state cost of finite-width ribbons per unit area
\begin{equation}\label{eq: e_sla}
\epsilon_{\rm sla}=C_0 J^{1/3}\sigma^{2/3}\varphi^{2/3}
\end{equation}
where $C_0=(9\pi^2/2)^{1/3}$.  Note that, when normalizing this cost relative to the characteristic energy density $J \varphi$, we find $\epsilon_{\rm sla}/(J \varphi) \propto (\ell_{\rm v}/\ell_{\rm c})^{2/3} = (\ell_{\rm d}/\ell_{\rm v})^2$, showing that the cost of finite domain formation is controlled by the ratio of the mesoscopic self-limiting width to inter-vortex spacing. Since, as we discuss next, the energy density of the optimal vortex array is roughly $J \varphi$, this suggests that finite-domain formation is energetically favorable compared to defective bulk states when self-limiting domain size is small compared to the characteristic defect spacing. This result is consistent with the intuitive argument put forward in ref.~\cite{HackneyPhysRevX.13.041010}.

\subsection{Bulk vortex lattice\label{sec: abrikosov}}

We next consider the energetics of the defect bulk phase.  As previously described by Hackney et. al.~\cite{HackneyPhysRevX.13.041010} and shown in Fig. \ref{fig: example simulations}, the bulk condensed state of the lattice frustration model forms a quasi-regular array of vortices with empty (i.e. voided) defect cores, leading to a quasi-uniform sponge morphology.  To understand the ground state energetics of this state, we first briefly consider the fully-occupied ($\Phi \to 1$) case, in which defect cores are completely filled in and can be modeled as an infinite plane tiled by a hexagonal Abrikosov lattice of like-charged vortices. Vortices are singular, non-analytic field configurations around which the phase field winds by integer multiples of $ 2 \pi$. We define the areal density of defects as
\begin{equation}
s(\mathbf{x})=\sum_{n,m} s_{nm} \delta(\mathbf{x}-\mathbf{x}_{nm})
\end{equation}
where $s_{nm}=\pm 2 \pi $ is the topological winding of each elementary vortex and $\mathbf{x}_{nm}$ is the location of the vortex on site $nm$ in the array,
\begin{equation}\label{eq: hexagonal lattice vectors}
\mathbf{x}_{nm}=d\Big(n+\frac{m}{2},\frac{\sqrt{3}}{2}m\Big)
\end{equation}
where $d$ is the nearest neighbor spacing of vortices on a triangular lattice and  $n,m\in[-\infty,\infty]$. Following a standard approach~\cite{nelson1995defects}, we can solve for the phase field produced by the superposition over vortices in the infinite array and compute the elastic energy.

To do so, we note that the intra-aggregate phase field can be related to the areal defect density field via Stoke's theorem:
\begin{equation} \label{eq: stokes}
\nabla \times (\nabla \theta) = s(\mathbf{x})
\end{equation}
where $\nabla \times {\bf v} = \epsilon_{ij} \partial_i v_j$ is the 2D curl of vector ${\bf v}$.  Following a standard approach~\cite{nelson1995defects}, we introduce the conjugate field $\tilde{\theta}$ defined by $\partial_i \theta= \epsilon_{ij} \partial_j \tilde{\theta}$ for which eq. (\ref{eq: stokes}) becomes an effective Poisson equation, $\nabla^2 \tilde{\theta} = s(\mathbf{x})$.  From these equations, we can solve for the elastic cost of the defect array and background frustration as
\begin{equation} \label{eq: wignersum}
\frac{J}{2}\int d^2 {\bf x}\big|\nabla \theta - {\bf A}\big|^2 = \frac{J A }{2} \sum_{{\bf G}\neq 0 } \frac{\rho_s^2}{|{\bf G}|^2}, 
\end{equation}
where ${\bf G}=4 \pi/(\sqrt{3} d) \big(\sqrt{3} h/2, k-h/2 \big)$ defines the reciprocal lattice vectors for integer values of $h$ and $k$. Here $\rho_s = 2 s/3 d^2$ is the areal topological charge density of the array, which must be set to $2\pi \varphi$ to neutralize the far-field  elastic cost of the background frustration by removing the divergent ${\bf G } \to 0$ term in the sum above.  Assuming that $s = 2 \pi$, this gives $d = \sqrt{2/(3 \varphi)} \propto \ell_v $.  Notably, the sum over reciprocal vectors in eq. (\ref{eq: wignersum}) diverges at large $|{\bf G }|$ and must be cutoff at some maximal wave vector $\Lambda_{\rm max} \approx 1/a$, resulting an a logarithmically divergent elastic cost as the microscopic lattice dimension vanishes,
\begin{equation}
    \sum_{{\bf G}\neq 0 } \frac{1}{|{\bf G}|^2} \approx \int_{1/d}^{\Lambda_{\rm max}} \frac{dq}{q} = \ln (d/a) .
\end{equation}
This illustrates the sensitivity of the bulk state to the {\it core size}, $r_c$, of the vortices and notably agrees with the observation that, in the bulk phase of the lattice GFA model, the vortex cores adopt a finite radius. Therefore, it is intuitive to understand that, for $\sigma/J \ll 1$, the core size will be much larger than the microscopic cutoff, $a$, due to the competition between the elastic energy, which favors larger cores, and the cohesive energy, which favors smaller cores.


To model the effect of the finite core on the bulk state,  we consider a hole of radius $r_{\rm c}$ around each vortex in the assembly.  To assess the optimal structure, we consider an approximate unit cell calculation of the phase-strain in which we model the hexagonal cell that surrounds each vortex in the triangular lattice by a circular one (see schematic in Fig. \ref{fig: hole schematic}b). In an infinite periodic vortex array, the elastic energy density, 
\begin{equation}
    \epsilon_{\rm elas}( \xv) \equiv \frac{J}{2}|\nabla\theta(\mathbf{x})-\mathbf{A}(\mathbf{x})|^2
\end{equation}
is invariant under symmetries of the arrangement.  In particular, normal gradients of the energy density vanish along the cell boundaries, which are lines of mirror symmetry in the triangular lattice, i.e. 
\begin{equation}
\label{eq: outerBC}
    (\mathbf{\hat{n}}\cdot\nabla)\epsilon_{\rm elas}( \mathbf{x}) \big|_{\partial {\cal C}} = 0
\end{equation} where $\partial C$ denotes the outer boundary of unit cell ${\cal C}$ and $\mathbf{\hat{n}}$ defines its normal.  
We approximate the unit cell as a circle of outer radius $R_0$ surrounding a single vortex at its center, which sits within a voided circular core of radius $r_c$ (see Figure \ref{fig: hole schematic}).  To determine the cell dimension, we superpose the phase strain generated by the central vortex 
\begin{equation}
\nabla\theta=\frac{s}{r}\hat{\phi}.
\end{equation} 
with the azimuthal gauge field $\mathbf{A}= \pi \varphi ~r\hat{\phi}$, to find the axisymmetric elastic energy density
\begin{equation}
    \epsilon_{\rm elas}( r) = \frac{J}{2} \Big(\frac{s}{r} - \pi \varphi r\Big)^2 .
\end{equation}
Applying the cell boundary condition in eq. (\ref{eq: outerBC}), $\partial_r \epsilon_{\rm elas}( R_0) =0$, 
results in the following condition of cell size,
\begin{equation}
    \pi R_0^2 ~\varphi = s. 
\end{equation}
This corresponds to the condition that the defect charges neutralize the background charge of the frustration field and additionally sets the inter-defect spacing $\ell_{\rm v} \approx 2 R_0 \propto \varphi^{-1/2}$.  

Given this optimal cell size (i.e. vortex density) we average the free energy density over the occupied annular region $r_c<r<R_0$ of a single cell, which be written as a function of reduced core size $\bar{r} = r_c/R_0$,
\begin{multline}
 \frac{\epsilon_{\rm defect}(\bar{r})}{J/R_0^2}=-s^2\frac{\ln \bar{r}}{1-\bar{r}^2}-\frac{ s}{2}(\varphi R_0^2) +\frac{1}{16}(\varphi R_0^2)^2(1+\bar{r}^2)
\\ +\frac{2}{1-\bar{r}^2}\frac{\bar{r}}{\bar{\lambda}}\\
\end{multline}

which we have written in dimensionless form by introducing the scaled variables $\bar{r}=r_{\rm c}/R_0$ and $\bar{\lambda}=(J/\sigma)R_0^{-1} \propto \ell_{\rm c}/\ell_{\rm v} \propto \varphi^{1/2}$.

\begin{figure*}[ht]
\includegraphics[width=0.9\textwidth]{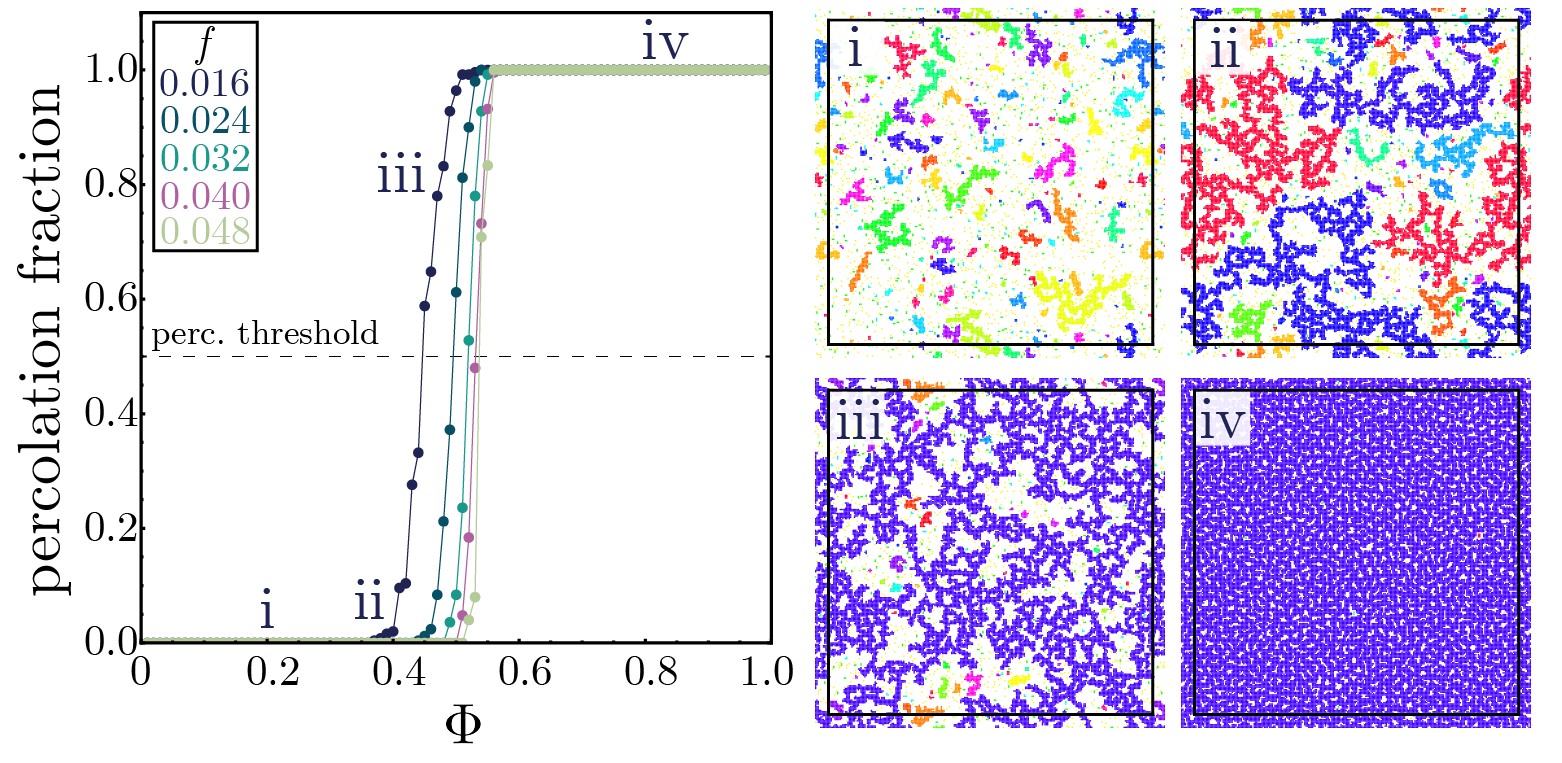}
\caption{Percolation of width-limited domains. Measured percolation fraction as a function of concentration shown for several different values of strong frustration. Dashed horizontal line denotes a percolation fraction of $0.5$. We define the percolation threshold, $\Phi_{\rm perc}$, to be the concentration at which the percolation fraction is equal to $0.5$. The labeled points i-iv correspond to simulation snapshots for a sequence of concentration values ($\Phi=0.2,0.4,0.5,0.8$) spanning the percolation transition. Here a small window on either side of the periodic boundary is shown with the simulation box denoted by the solid black square. Each aggregate is assigned a unique color to highlight the emergence of a system spanning cluster as concentration is increased. All simulations were run with the following parameters: $\Sigma/J=0.09$, $\beta J=40$ and $L=250$.}\label{fig: percolation fraction}
\end{figure*}

Observing that $\varphi R_0^2=2s$, we can minimize $\epsilon(\bar{r})$ with respect to $\bar{r}$ to obtain a transcendental equation for the optimal dimensionless hole size $\bar{r}_*$:

\begin{equation}\label{eq: lambda}
\bar{\lambda}s^2=\frac{1+\bar{r}^2}{\frac{1}{\bar{r}}(1-\bar{r}^2)^2+2\bar{r}\ln\bar{r}-\frac{\bar{r}}{2}(1-\bar{r})^2}
\end{equation}
which can be approximately satisfied in the limit of small and large $\bar{\lambda}$ as:
\begin{equation}
\bar{r}_*\simeq
\begin{cases}
\bar{\lambda} & \bar{\lambda}\rightarrow 0 \\
1-\bar{\lambda}^{-1/3} & \bar{\lambda}\rightarrow \infty
\end{cases}
\end{equation}
Plugging this back into the expression for $\epsilon(\bar{r})$, we find the limiting form of the optimal sponge energy density:
\begin{equation}\label{eq: e bulk}
\frac{\epsilon(\bar{r}_*)}{J\varphi}=
\begin{cases}
\ln\big(\bar{\lambda}\big) & \bar{\lambda}\rightarrow 0  \\
\bar{\lambda}^{-2/3} &\bar{\lambda}\rightarrow \infty
\end{cases}
\end{equation}
Similarly, we can approximate the optimal hole concentration, $\Phi_{\rm hole}=\bar{r}_*^2$, as:
\begin{equation}\label{eq: phi hole}
\Phi_{\rm hole}\simeq
\begin{cases}
\bar{\lambda}^2 & \bar{\lambda}\rightarrow 0  \\
1-2\bar{\lambda}^{-1/3} &\bar{\lambda}\rightarrow \infty
\end{cases} .
\end{equation}
In the limit of low frustration, where $\bar{\lambda} \propto \varphi^{1/2} \to 0$, the optimal hole size goes to a constant value of $r_* \approx \ell_{\rm c}$ set by the elasto-cohesion scale such that the optimal vortex density increases as the inter-defect spacing diverges in the zero frustration limit.  In the large frustration (or low cohesion when $\ell_{\rm c} \gg  \ell_{\rm v}$) limit, optimal hole size approaches the inter-vortex spacing, but with a finite inter-vortex ``gap'' $2(R_0-r_*) \approx R_0 \bar{\lambda}^{-1/3} \propto \ell_{\rm v}^{2/3}\ell_{\rm c}^{1/3} \sim \varphi^{-1/3}$, which is distinct from (i.e. larger than) the self-limiting domain size scaling.

For low-enough temperature, we assume that the relationship between hole size and concentration derived above will hold for both the uniform vortex sponge at high concentration (when the total density exceeds $1-\Phi_{\rm hole}$)  as well as the phase-separated condensates found at weak frustration. Thus, we use this predicted dependence of vortex core size and inter-vortex spacing as a basis for comparing the structural similarity of these two types of bulk states.

\section{Concentration Driven Transition from Self-Limiting to Bulk at Strong Frustration}\label{section: conentration}

In section \ref{section: intro}, we suggested that the dilute state of dispersed/self-limiting assembly, found at strong frustration, ultimately evolves into a macroscopically large bulk phase upon increasing concentration via a percolation of finite width domains, first into a heterogneous network of finite width domains and then ultimately into a relatively ordered bulk array of defect-enclosing holes at higher concentration. We investigate this evolution with MC simulations of the lattice model described in section \ref{section: minimal model} by running simulations at fixed $f$, $\Sigma/J$, $\beta J$, and increasing concentration, $\Phi$. In particular, we analyze the statistics of clusters of assembled particles, identified as connected graphs of occupied nearest neighbor sites on the square lattice.  

\begin{figure*}[ht!]
\centering
\includegraphics[width=\textwidth]{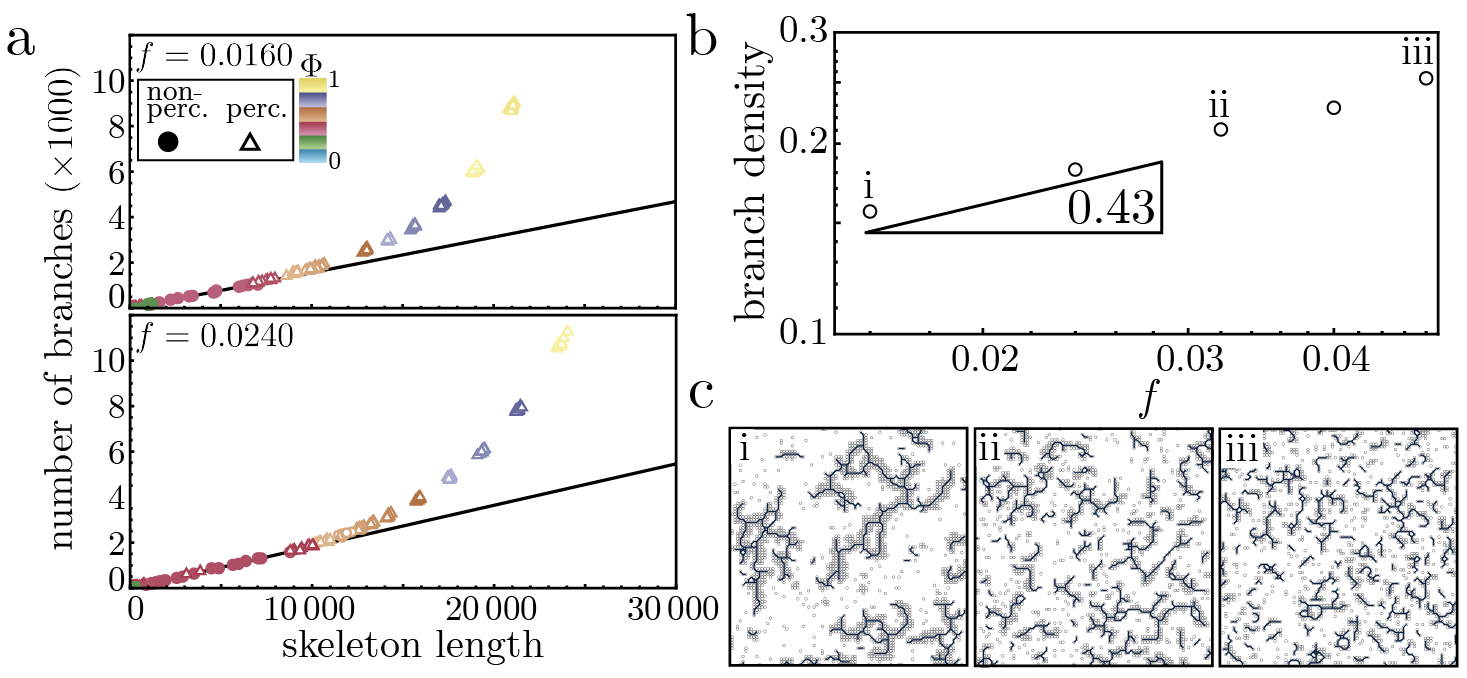}
\caption{Branch density in frustrated assembly. (\textbf{a}) The number of branch points along the morphological skeleton of our simulated aggregates is plotted as a function of the skeleton length. Each point corresponds to a single aggregate observed in simulation with filled circles and open triangles corresponding to non-percolating and percolating aggregates, respectively. Each point is colored according to the total concentration of the simulation in which it was observed. Results for two different values of frustration are given and the solid line corresponds to a linear least squares regression fit for each data set. The slope of these lines corresponds to the branch density. (\textbf{b}) Plots of the measured branch density as a function of frustration. From this, we observe that the branch density scales like $\sim \varphi^{2/5}$. (\textbf{c}) Simulation snapshots with the morphological skeleton of each aggregate highlighted in black. The snapshots labeled i,~ii,~iii correspond to frustration values of $f=0.016,0.032$ and $0.048$, respectively. All snapshots taken from simulations with $\Phi=0.3$, $\Sigma/J=0.09$, $\beta J=40$ and $L=250$. }\label{fig: branch density}
\end{figure*}

\subsection{Percolated networks of finite width domains}\label{section: percolation}

The percolation fraction of a given set of conditions is obtained by observing aggregated clusters at regular intervals and counting the fraction of time-steps containing a system spanning aggregate~\cite{stauffer2018introduction}. The percolation fraction as a function of concentration is plotted in Figure \ref{fig: percolation fraction} for several different values of frustration. Here, we find that the percolation fraction rapidly grows from zero at low concentration to near unity as $\Phi\rightarrow 1$. By defining the percolation threshold, $\Phi_{\rm perc}$, as the concentration at which a spanning cluster is observed half of the time, we quantitatively evaluate this transition. Doing so, we observe $\Phi_{\rm perc}\in[0.4,0.65]$, with the percolation threshold shifting to higher concentration with increased frustration where the self-limiting domains tend to become more narrow in width. This frustration dependence illustrates that the nature of the percolation transition in networks of self-limiting domains is different from the simplest classes of bond or site percolation on a square lattice, which have a fixed percolation threshold of $0.5$ and $\sim0.59$, respectively~\cite{derrida1985corrections}.  Lastly, we note that the tendency of increased frustration to shift the percolation threshold to higher concentration is consistent with findings of recent study of a similar lattice frustration model that considers phase behavior in the grand canonical ensemble~\cite{ortiz2024statistical}.


One likely factor influencing the shift in percolation threshold could be the tendency of frustration to alter the density of branch points in the aggregates, as branchy aggregates are more spatially dense and require more subunits (i.e. higher concentration) to span the lattice~\cite{zimm1949dimensions}. In Figure \ref{fig: branch density}a, we investigate this idea by counting the number of branches per aggregate as a function of morphological skeleton length (i.e. the length of 1D ``medial backbone'' at the center of the domains) for a range of different concentrations and frustrations. We observe that, in non-percolating aggregates, the number of branches grows roughly linearly with skeleton length, suggesting a uniform branch density below the percolation threshold, or equivalently the mean linear distance between branches is roughly constant and independent of $\Phi$. Furthermore, in Figure \ref{fig: branch density}b, we show that this branch density does indeed increase with frustration, albeit weakly. Going further, we observe that, near and above the percolation threshold, the number of branches grows super-linearly with skeleton length. In this regime, aggregates remain network-like, but realize a branch density that increases with concentration.  This suggests a transition in branching morphology of quasi-1D width-limited aggregates that occurs near the onset of percolation.

\begin{figure*}[ht]
\includegraphics[width=\textwidth]{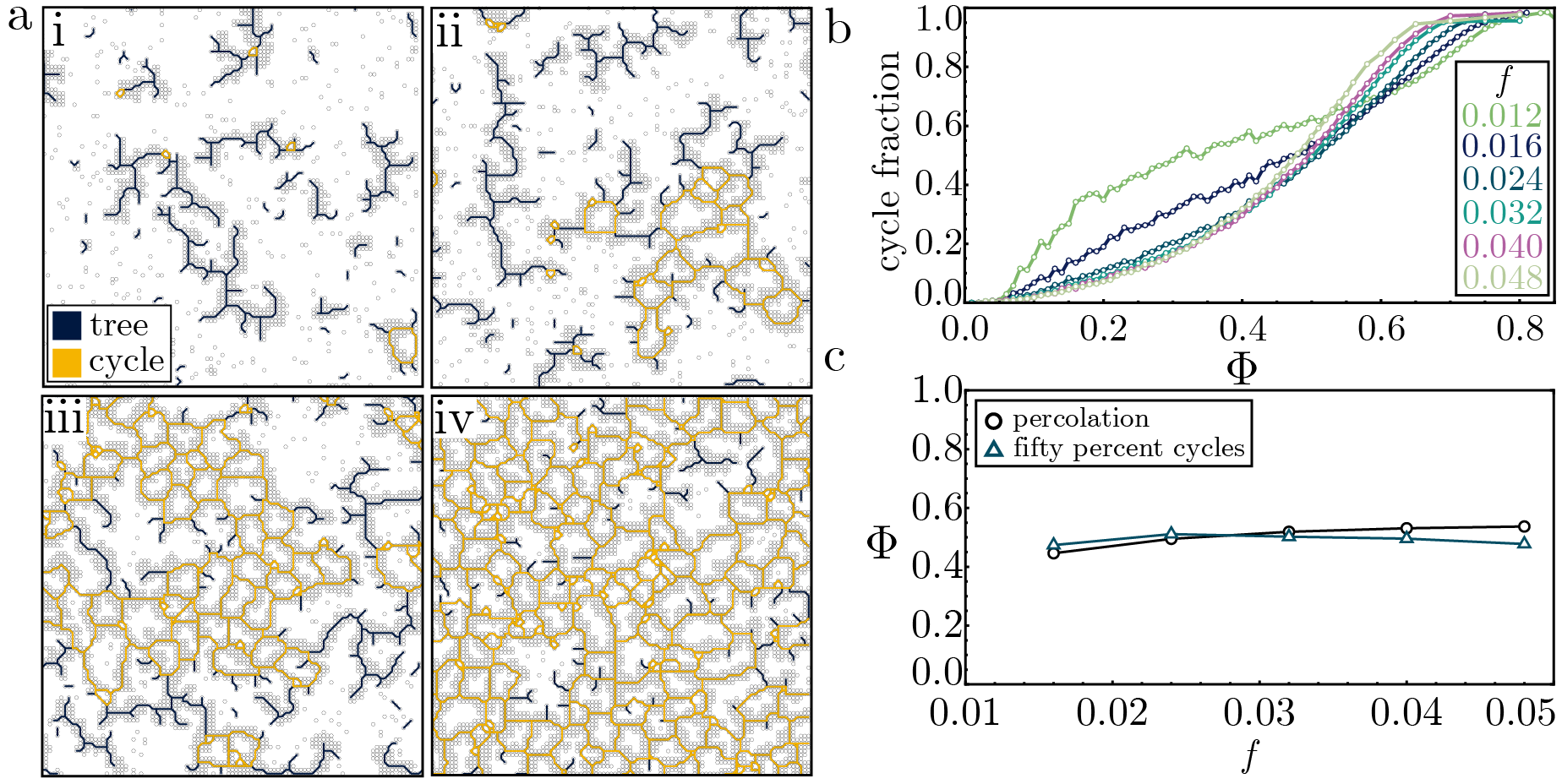}
\caption{Concentration 
driven transition in aggregate topology (\textbf{a}) i-iv Simulation snapshots for a sequence of concentrations ($\Phi=0.2,0.3,0.5,0.7$) spanning the percolation transition. The morphological skeleton of each aggregate is illustrated as a graph of edges connecting neighboring particles. Edges that belong to cycles below a certain length threshold are colored yellow and edges that belong to tree-like sections are colored navy blue. The cycle length threshold is chosen to be one and a half times the average cycle length containing a charge one defect in the optimal sponge phase. All simulations are run with $f=0.016$, $\Sigma/J=0.09$, $\beta J=40$ and $L=250$. (\textbf{b}) Fraction of morphological skeleton contained within a cycle. (\textbf{c}) Comparison between the cycle and percolation threshold as a function of frustration.}\label{fig: cycles}
\end{figure*}

\subsection{Tree-like versus loopy aggregate topology}\label{section: topology}

Careful analysis of aggregates reveals that this transition to super-linear growth in branch number can be associated with a rapid accumulation of loops of a characteristic dimension set by the preferred vortex size. We characterize this transition in aggregate topology via the fraction of the morphological skeleton contained within tree-like versus cyclic sections. In Figure \ref{fig: cycles}a, we show a sequence of states spanning the transition with the morphological skeleton colored to show the existence of cycles~\footnote{When counting looping spans of the graph, we employ an maximal cycle length corresponding to one and a half times the average cycle length measured in the optimal sponge phase corresponding to the energy density minima found in Figure \ref{fig: energy density}, since the optimal vortex size sets a preferred size of loops}. Here we see that, at low concentration, the system is dominated by tree-like aggregates with uniform branch density. At high concentration, the system is dominated by defect containing loops in the morphological skeleton. In between these two limits, we observe coexistence of dense clusters of locally loopy (defective) and tree-like (finite-width) aggregates. This crossover from trees to cycles is neatly summarized in Figure \ref{fig: cycles}b, where we show the fraction of morphological skeleton contained within cycles as a function of concentration. Interestingly we find that, at low concentration (but above the pseudo-critical aggregation concentration threshold) the cycle fraction increases as frustration is lowered; this suggests that the aggregates become less tree-like as the weak frustration regime (i.e the phase separation binodal) is approached. Conversely, at high concentration, we find that the cycle fraction increases with frustration, consistent with the increased density (i.e. smaller size) of defect holes. Furthermore, by defining the cycle threshold as the concentration where half of the morphological skeleton is contained within loops that surround defect holes, we can compare the location of this transition in aggregate topology to the formation of a percolating network (see Figure \ref{fig: cycles}c). From this, we see that these two tendencies coincide and that the onset of percolation\textemdash which is a global phenomena and is only well defined in the thermodynamic limit\textemdash is strongly correlated with the structural evolution of aggregated clusters from predominately tree-like clusters into loopy networks.


\subsection{Evolution from heterogeneous networks to quasi-uniform bulk ``sponge"}\label{section: topology}

We next analyze the correlations between defect enclosing loops in the assembled aggregates and their dependence on concentration.  Near the onset of percolation, coexistence of the two populations of aggregate morphology leads to a state characterized by dense clusters of defect holes connected by tree-like ribbons\textemdash resulting in a bulk phase with a large degree of non-uniformity. This is distinct from the high concentration bulk phase, where the assemblies are more uniformly defect-riddled in terms of hole size and spacing. We describe these distinct structures observed at concentrations near and far above the percolation/cycle threshold as {\it heterogeneous networks} and {\it (quasi-)uniform defect sponges}, respectively. Additionally, we note the similarity between the structure of the uniform defect sponge and the bulk aggregates observed at weak frustration (i.e. uniformly spaced holes of a given characteristic size) suggesting that the condensed phase represents phase separation between the low concentration monomer gas and the high concentration uniform defect sponge phase.



\begin{figure}[]
\centering
\includegraphics[width=0.5\textwidth]{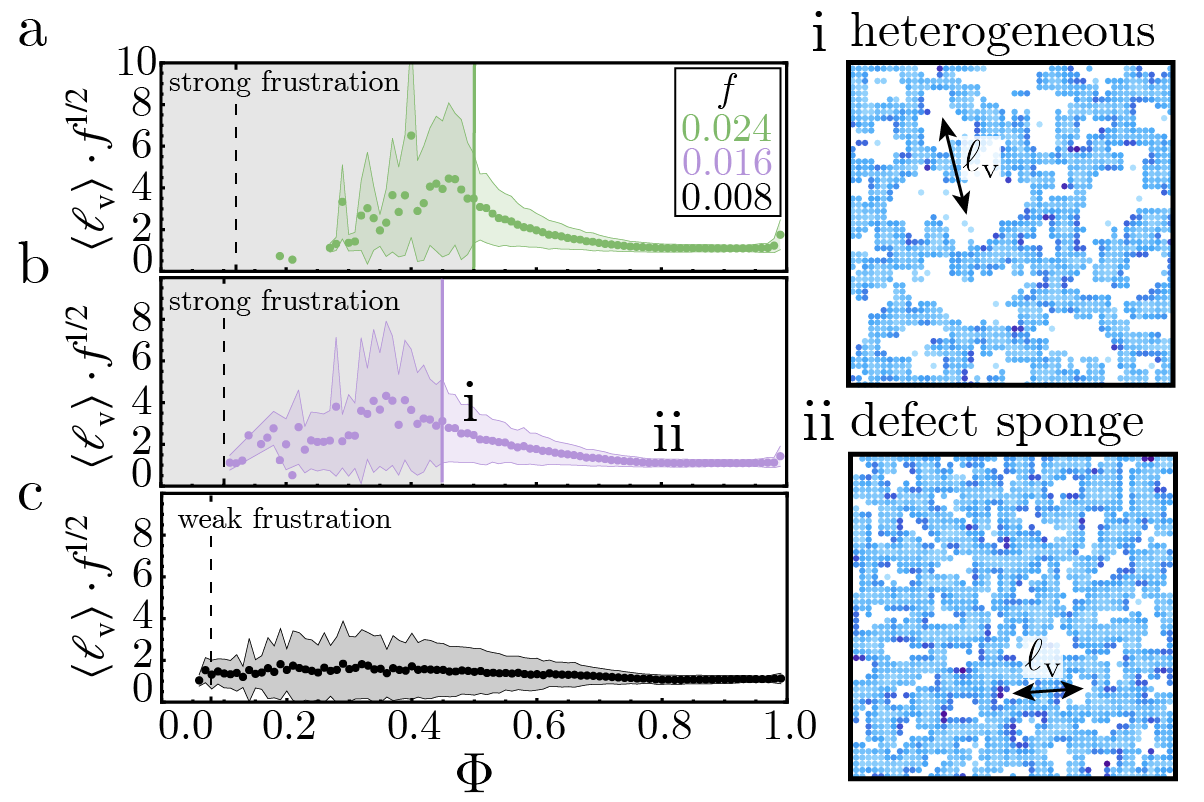}
\caption{Hole spacing at strong and weak frustration. (\textbf{abc}) Average inter-hole spacing, $\langle \ell_{\rm v}\rangle$, normalized by the expected defect spacing, $f^{-1/2}$. Distance calculated with respect to the geometric center of each hole. The colored dots correspond to the average hole spacing and the shaded region around the dots denote the variance. The vertical solid lines denote the location of the percolation transition occurring as a function of concentration at strong frustration. The vertical dashed lines denote the critical aggregation concentration. The region below the percolation transition is shaded in gray to emphasize that this region does not contain a system spanning aggregate. Numbered points i and ii correspond to the simulation snapshots depicting the heterogeneous and defect sponge phases, respectively. All simulations run with the parameters: $\Sigma/J=0.09$, $\beta J=40$ and $L=250$.}\label{fig: hole size and spacing}
\end{figure}

The distinction between these states can be seen by looking at the average spacing between holes, normalized by the expected center-to-center spacing between defects (i.e. $\ell_{\rm v}\sim \varphi^{-1/2}$) in the fully occupied Abrikosov vortex lattice, shown in Fig.~\ref{fig: hole size and spacing}. At weak frustration, the normalized hole spacing is a constant of order one across the entire range of concentration (Fig.~\ref{fig: hole size and spacing}c), consistent with the expected Abrikosov-like structure of the bulk condensates. Above the binodal (Fig.~\ref{fig: hole size and spacing}a-b), the hole spacing rises from zero to a peak at concentrations near to the percolation threshold. When $\Phi\gg\Phi_{\rm perc}$, the normalized spacing of the strongly frustrated aggregate continuously drops to $\sim 1$. Thus indicating that the structure of the percolated aggregates gradually converges to the Abrikosov-like structure of the phase separated condensates, consistent with the emergence of a space filling network of uniformly sized loops enclosing topological defects. Lastly, we observe a sharp increase in the inter-hole spacing in the limit that $\Phi\sim 1$. This can be easily understood as the limit where the holes start to completely fill in, leading to a divergent spacing between a tiny number of empty cores.

The agreement between the center-to-center hole spacing and the predicted inter-defect spacing suggests that the intra-assembly holes might also exhibit inter-vortex order (e.g. trianglar lattice) at least at sufficiently low temperature. However, despite the regularity in spacing, the observed array of holes don't display a large amount of orientational order (see Appendix \ref{appendix: orientational order}). This suggests that the temperature range in the vicinity of the phase separation binodal is above the melting point of the Abrikosov defect crystal, at least for the conditions simulated in this study. This defect melting transition has been studied both theoretically~\cite{fisher1980flux,doniach1979topological} and numerically~\cite{franz1995vortex,wheatley1995flux} in the (fully occupied) frustrated 2D XY model. Franz and Teitel report a defect melting transition occurring at $T_{\rm m}\simeq 0.007 J$~\cite{franz1995vortex}, which is much cooler than the lowest temperature of $T=0.014J$ studied here. 

\begin{figure*}[ht]
\centering
\includegraphics[width=\textwidth]{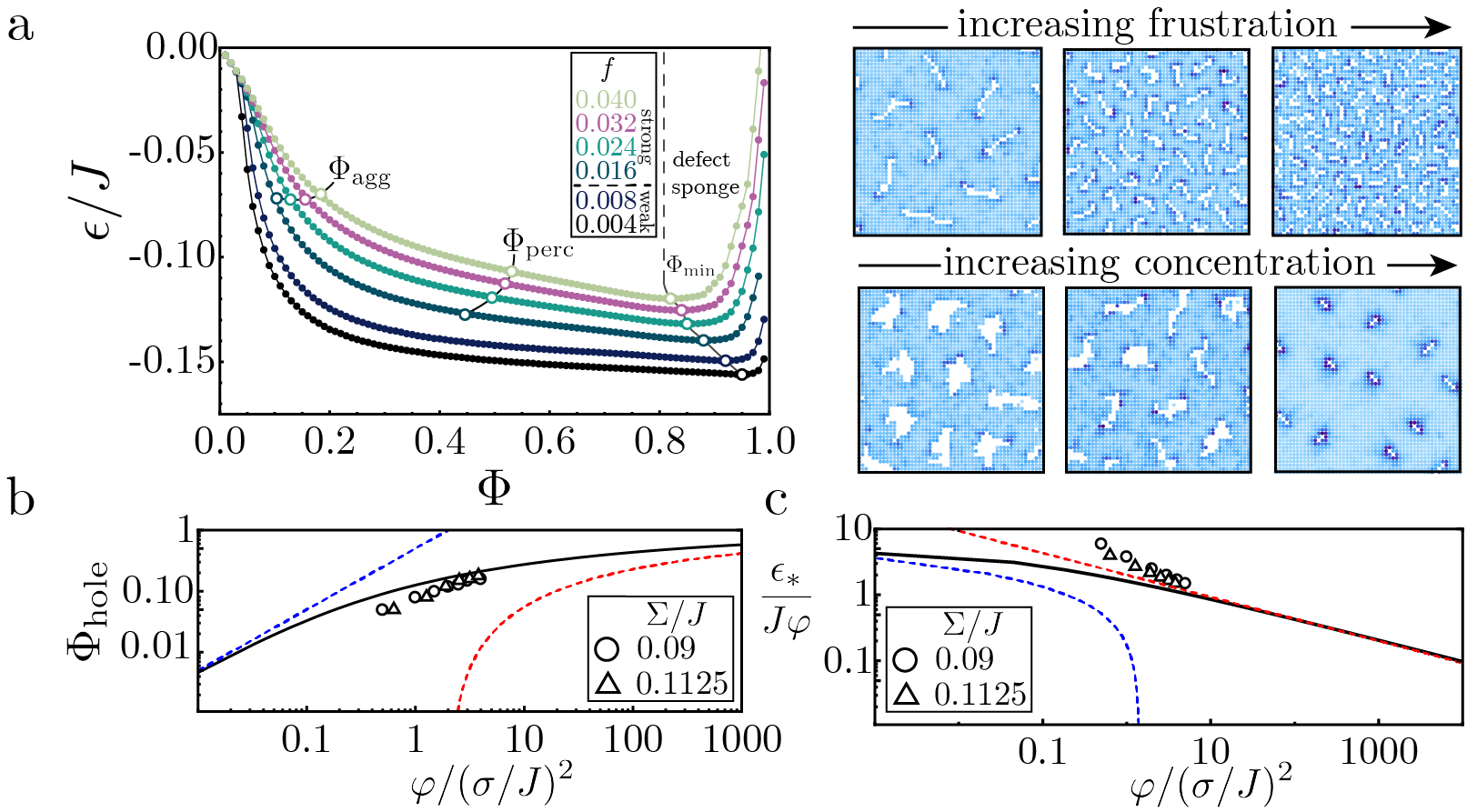}
\caption{Energy per subunit across concentration range. (\textbf{a}) The energy per subunit measured from numerical simulations that sweep over the entire range of concentration is shown for several values of frustration ranging from strong to weak. The location of the minimum corresponds to the concentration, $\Phi_{\rm min}$, at which the optimally dense sponge phase extends across the entire system. The location of the percolation and aggregation concentrations are also marked for clarity. (\textbf{b}) Concentration of intra-aggregate holes (defined as $\Phi_{\rm hole}=1-\Phi_{\rm min}$) as function of the ratio between frustration and the cohesion-to-stiffness ratio squared. Plot markers correspond to simulation data with $\Sigma/J=0.09$ (open circles) and 0.1125 (open triangles). The dotted blue and red lines denote the limiting behavior (see eq. (\ref{eq: phi hole})) as $\varphi/(\sigma/J)^2\rightarrow 0$ and $\infty$, respectively. (\textbf{c}) Comparison between the predicted bulk energy density (see eq. (\ref{eq: e bulk})) and the bulk energy density measured in simulation. The dotted blue and red lines denote the limiting behavior as $\varphi/(\sigma/J)^2\rightarrow 0$ and $\infty$, respectively. Simulation snapshots for sequence of increasing frustration and increasing concentration illustrate the effect that varying these parameters has on equilibrium size and density of defect holes. Here the snapshots show a small subset of entire simulation, centered on sponge interior, so the effect can clearly be seen.}\label{fig: energy density}
\end{figure*}

In Figure \ref{fig: energy density}a, we show the energy per subunit as a function of increasing concentration. At strong frustration, we see that there is a well defined global minimum at some critical concentration, $\Phi_{\rm min}$. The location of this minimum corresponds to the concentration at which the sponge phase can uniformly achieve the preferred hole size and density over the entire lattice.  That is, at this particular density, $\Phi_{\rm min}$, the space-filling vortex sponge achieves both the selected spacing and core size of vortices described in Sec.~\ref{sec: abrikosov}.  Above this point, the holes are forced to fill in to a higher density state leading to additional phase strain around the defect core and an increased energy per subunit. Conversely, at concentrations lower than $\Phi_{\rm min}$, there is an excess of free space that can be filled in for an energetic gain. This excess space results in a larger number of conformational degrees of freedom, leading to an increase in structural heterogeneity. Hence, we use the location of the minimum of energy density as function of concentration as a means of delineating the boundary between the heterogenous network and quasi-uniform vortex sponge.

In section \ref{sec: abrikosov}, we introduced the circular cell approximation to describe the energetics of the homogeneous sponge phase. This simplified model predicts that the sponge phase should have an optimal hole density, $\Phi_{\rm hole}$, controlled by the ratio of $\varphi/(\sigma/J)^2$ and can be obtained from the transcendental equation for the optimal hole size given in eq. (\ref{eq: lambda}). In the limit that $\varphi\ll(\sigma/J)^2$ (which corresponds to the zero-temperature definition of weak frustration) we find that $\Phi_{\rm hole}\simeq\varphi/(\sigma/J)^2$. Conversely, in the limit that $\varphi\gg(\sigma/J)^2$, $\Phi_{\rm hole}$ can be expressed as a polynomial of $\varphi/(\sigma/J)^2$ (see eq. (\ref{eq: phi hole})).
This can be related to the concentration that minimizes the per subunit aggregation energy via the relation:
\begin{equation}
\Phi_{\rm hole}=1-\Phi_{\rm min}.
\end{equation}
In Figure \ref{fig: energy density}b, we test this prediction for the dependence of hole density on frustration and the cohesion-to-stiffness ratio. Comparing our simulation results to the two asymptotic limits, as well as to the full non-linear solution of eq. (\ref{eq: lambda}), we find remarkably good agreement with our simplified model of optimal bulk vortex sponges.  This comparison confirms the basic trend that $\Phi_{\rm hole}$ increases with frustration and, more specifically, shows that our simulation parameters necessarily fall in the crossover range between the two asymptotic regimes, as this is the regime where self-limiting assembly occurs. Similarly, the dimensionless energy density, $\epsilon/J\varphi$, of the optimal sponge phase can also be obtained from the solution to eq. (\ref{eq: lambda}). Comparing this result to our simulation data (Figure \ref{fig: energy density}c) we again find good qualitative agreement between simulation and theory.


Lastly, we observe from Fig. \ref{fig: energy density}a that the energy density landscape near the minimum becomes less convex as frustration is decreased and is essentially flat in the limit of weak frustration. This nearly linear behavior is consistent with the coexistence between a dispersed monomer gas and a single condensed defect bulk observed below the binodal. Conversely, the results described in sec.~\ref{section: topology} suggest that the coexistence at high frustration is more complex, consisting of phase separation between tree-like self-limiting aggregates of finite width and localized clusters of vortices that can at least partially condense on the self-limiting aggregates themselves.


Throughout this section, we saw that the dispersed/SLA phase found at dilute concentration and strong frustration undergoes a percolation transition to a bulk sponge phase at intermediate concentration. Near the percolation threshold, the internal structure is heterogeneous and marked by the presence of loops and branches of highly variable size. However, the structure becomes more uniform with increasing concentration and gradually approaches the quasi-uniform bulk state. The internal structure of this state is similar to the Abrikosov lattice found in the uniformly frustrated XY model, only with a punctured disc surrounding each defect core, with relatively uniform size for fixed values of $\varphi$ and $\sigma/J$. Additionally, we saw that the structure of this high concentration assembly is locally similar to that of the topologically defective condensate found at weak frustration, confirming that this latter phase represents coexistence between the low concentration dispersed phase and the high concentration homogeneous sponge. In the following section, we turn our attention to this phase separated state and the nature of the transition from self-limiting to bulk condensation at low to intermediate concentration driven by decreasing frustration.

\section{Frustration Driven Transition from Self-limiting to Bulk Condensation}\label{section: condensation}

Below the percolation concentration, assembly of the lattice GFA model exhibits a state of self-limiting aggregation above a critical concentration and, below that, bulk phase separation into vapor and bulk, defect-sponge states. Previously, we found that the critical frustration $\varphi_c$ increased with the ratio of cohesion to stiffness ratio as would be expected from zero-temperature arguments~\cite{HackneyPhysRevX.13.041010}, but also exhibited a temperature sensitivity that cannot be explained by energetics alone.  In this section, we consider thermodynamic effects of finite temperature and their impact on the frustration-driven transition between these states.

The phase boundary separating the self-limiting and condensed states of assembly can be described as the set of conditions where the difference in free energy density between these two states is zero. Thus, we can write down a simple equation for the phase boundary as:
\begin{equation}\label{eq: phase boundary}
\frac{\Delta F}{A}=\epsilon_{\rm sla}-\epsilon_{\rm bulk}-T\frac{\Delta s}{a^2}=0
\end{equation}
where $\epsilon_{\rm sla}= C_0 J^{1/3}\sigma^{2/3}\varphi^{2/3}$ is the excess energy density of the self-limiting domains (see eq. (\ref{eq: e_sla})) and $\epsilon_{\rm bulk}\simeq C_1 J\varphi$ is the excess energy density of the areal (holey) vortex array. We define $\Delta s$ as the entropy density difference between these two states (i.e. the specific entropy change). Here it is important to note that, while the prefactor $C_1$ is a constant in the limit of vanishing defect core size (see eq. (\ref{eq: wignersum})), it will generically have some dependence on the dimensionless ratio $\varphi/(\sigma/J)^2$ which selects the optimal hole size.  However, as our circular cell approximation suggests that $C_1$ is only weakly dependent on $\varphi/(\sigma/J)^2$ (see eq. (\ref{eq: e bulk})), we will assume a constant value of $C_1\approx 1$. This is in agreement with the results for the dimensionless bulk energy depicted in Figure \ref{fig: energy density}c. When $T=0$, the entropy difference between these states can be neglected and this equation can be solved to find the critical frustration separating these two states:
\begin{equation}
\varphi_{\rm c}=\bigg(\frac{\sigma}{J}\bigg)^2
\end{equation}
with $\varphi>\varphi_{\rm c}$ (i.e. \textit{strong frustration}) corresponding to the self-limiting state and $\varphi<\varphi_{\rm c}$ (i.e. \textit{weak frustration}) corresponding to the phase separated state. Note that this criterion for $\varphi/\varphi_{\rm c}$ greater than or smaller than unity is equivalent to consideration of the ratio of optimal vortex spacing to finite domain thickness, $\ell_{\rm v}/\ell_{\rm d}$, originally derived in Reference~\cite{HackneyPhysRevX.13.041010}.  Initial comparison of this scaling prediction to simulation results extracted from the lattice GFA model suggested, at best, a substantially lower power-law dependence of $\varphi_{\rm c}$ on $\sigma/J$ and an unpredicted shift with temperature to lower frustration. Although surprising, consideration of this result as the zero-temperature limit of equation (\ref{eq: phase boundary}) suggests the important role that entropy plays in stabilizing the self-limiting phase of assembly relative to an unlimited bulk state. Indeed, this scenario agrees with the result of a recently studied chain model of frustrated assembly~\cite{wang2024thermal} that suggests that finite temperature\textemdash at least in one dimensions\textemdash is a necessary condition for self-limiting assembly. Similarly, it suggests that thermal effects that couple to the relative entropy of assembly, ignored by considerations of the ground state energy alone, are strong under conditions where the lattice GFA model exhibits self-limiting assembly.

While the entropy differences between self-limiting and bulk states of assembly can in principle be computed by thermodynamic integration~\cite{frenkel2023understanding}, it not necessarily possible to attribute those quantities to the microscopic origins of their underlying fluctuations. Instead, we analyze several different sources of fluctuations that can distinguish between the overall specific entropy of these two states of assembly.  Based on this comparison, we propose a simplistic mean-field type theory for the entropic free energy difference between self-limiting and bulk condensed states, and use it to understand the nature and magnitudes of finite-temperature effects on the critical frustration.


\begin{figure*}[ht!]
\centering
\includegraphics[width=\textwidth]{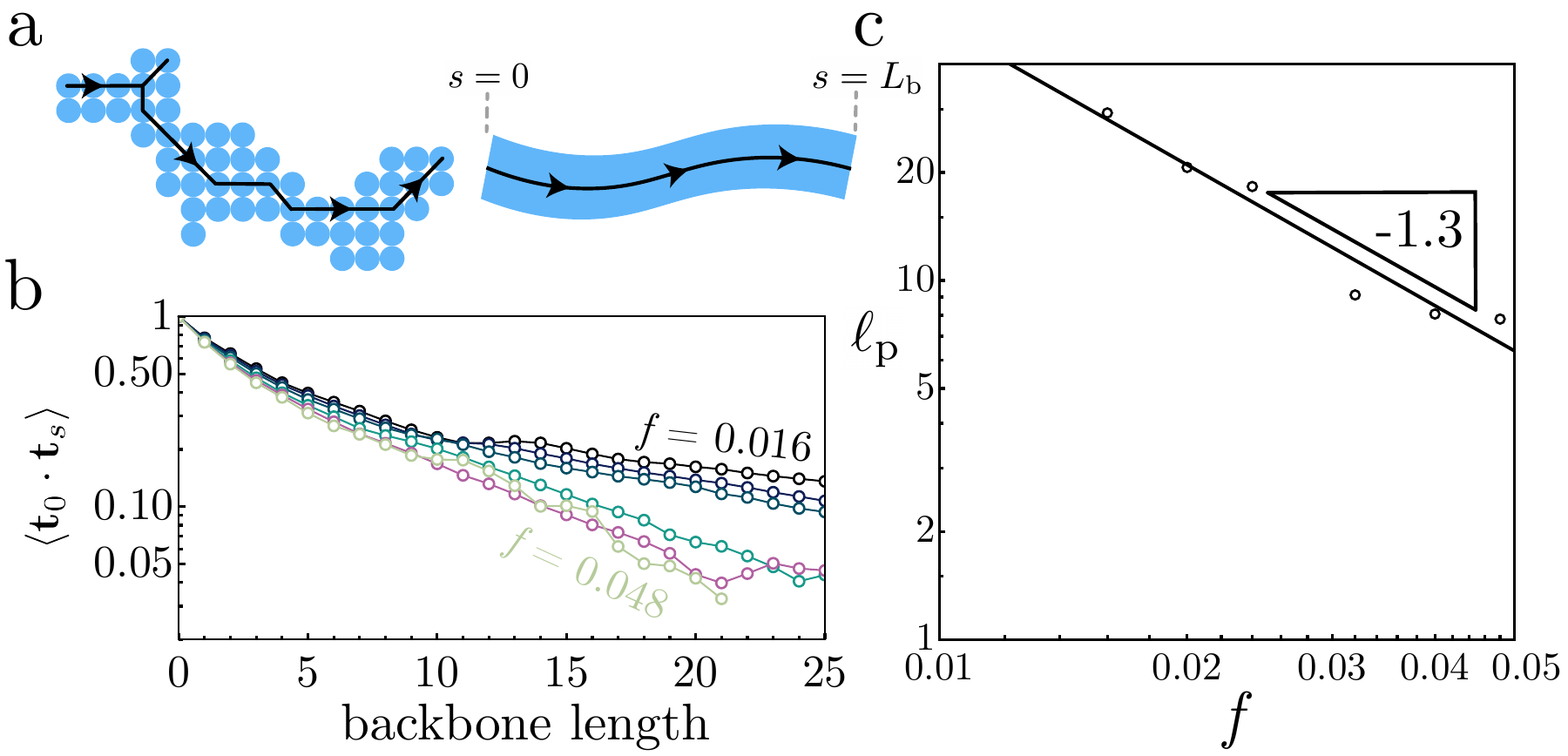}
\caption{Persistence length of self-limiting aggregates. An example of both a simulated and continuum theory aggregate is shown in (\textbf{a}) with the morphological skeleton denoted via a black line. Arrows are drawn to highlight the variation in tangent vector along the skeleton. The persistence length is obtained by fitting to (\textbf{b}) the long range tangent-tangent correlation function measured over a range of frustration. The short range scaling behavior is an artifact  of discrete lattice effects and intra-assembly branching that disappears upon coarse graining the aggregate backbone (see appendix \ref{appendix: bending entropy}). (\textbf{c}) Persistence length of self-limiting aggregates plotted as function of frustration. Triangle denotes the slope of the least squares regression, which agrees with the expected scaling (see eq. (\ref{eq: bending modulus})). All simulations run with $\Sigma/J=0.09$, $\beta J=40$ and $L=250$.}\label{fig: persistence length}
\end{figure*}

\subsection{Conformational entropy}

Conformational entropy refers to the entropy associated with fluctuations in aggregate shape and internal order, a property assembled aggregates~\cite{israelachvili_intermolecular_1992} have in common with macromolecular objects~\cite{grosberg1995statistical}. In the lattice GFA model, aggregates can take on a variety of different conformations that deviate from their locally favored ground state morphology by bending, branching, looping or changing width via capillary (edge) fluctuations. Additionally, the individual subunits in our model have an internal orientational degree of freedom that can fluctuate around its preferred arrangement, which constitute effective ``spin wave'' fluctuations around the ground state order. 

While all of these conformational fluctuations have an associated entropy that contributes to the total of each phase, they do not necessarily distinguish between the two states thermodynamically. In particular, we argue that the specific entropy associated with spin fluctuations and capillary edge fluctuations are not likely to be very different in self-limiting and bulk (vortex sponge) states.  In Appendix \ref{appendix: orientational entropy} we show that phase-fluctuations are largely governed by $\beta J$ and are fairly insensitive to frustration as well as large scale morphological differences between bulk sponge and self-limiting domains.  Additionally, in Appendix \ref{appendix: capillary fluctuations} we compare the capillary edge fluctuations between the bulk and self limiting states.  Since the relatively large core size of the vortex sponge permits a comparable range of edge fluctuations to the self-limiting state, we find that free edge fluctuations of both bulk and self-limiting states are well described by a simple model for capillary fluctuations along a two-dimensional ribbon.  The quantitative similarity of these two types of conformational fluctuations suggests that they are unlikely to contribute significantly to the free energy difference between bulk and self-limiting states.


Unlike edge and phase fluctuations, there is a more drastic difference between the 2D paths of the backbones of self-limiting ribbons (i.e. fluctuating ``worm-like'' ribbons) and bulk sponges (i.e. organized arrays of holes). Here we consider the conformational entropy associated with the backbone of self-limiting domains, akin to the bending fluctuations of a polymer~\cite{khokhlov1994statistical}. In the condensed phase, these fluctuations are highly-constrained by the energetic preference for closed loops of a given size and spacing. Therefore, to a first approximation, we estimate that there is no entropy of backbone domain fluctuations in the bulk phase.  

We consider a finite-width ($W^*$) ribbon domain to have the effective bending elastic free energy of worm-like polymer~\cite{Marko1995}:
\begin{equation}\label{eq: bending energy}
H_{\rm bend}=\frac{B}{2}\int\bigg(\frac{\partial^2 \mathbf{r}}{\partial s^2}\bigg)^2\rm ds
\end{equation}
where $\mathbf{r}(s)$ is the trajectory along the backbone as a function of arc length $s$, $\partial^2 \mathbf{r}/\partial s^2$ is the instantaneous backbone curvature and $B$ is the bending stiffness per unit length. In the context of our model, the bending stiffness of the self-limiting aggregates can be written as:
\begin{equation}\label{eq: bending modulus}
B=\frac{J\pi^2\varphi^2}{72}W_{*}^5=\frac{J\pi^2}{72}\bigg(\frac{\sigma}{J}\bigg)^{5/3}\varphi^{-4/3}
\end{equation}
where $W_*\simeq\ell_{\rm d}=(\sigma/J)^{1/3}\varphi^{-2/3}$ is the preferred self-limiting width.  This form derives from considering inverse square radius terms of the excess energy density of finite-width annular domains of the continuum model in the limit of large radius (see Appendix \ref{appendix: bending energy}).  Using this, we can evaluate the effect of bending stiffness on conformational fluctuations by calculating the tangent-tangent correlation:
\begin{equation}
    \langle\mathbf{t}_i\cdot\mathbf{t}_j\rangle\simeq e^{-s/\ell_{\rm p}}
\end{equation}
where $\mathbf{t}=\partial\mathbf{r}/\partial s$ is a vector tangent to the aggregate backbone and $\ell_{\rm p}=\beta B$ defines the persistence length beyond which bend deformations become important. The tangent-tangent correlation measured along the backbone of our self-limited aggregates is given in Figure \ref{fig: persistence length}. From this, we observed two different scaling regimes. The first of which is a short range ($s\lesssim 10$) regime where the correlation length has a weaker dependence on frustration than that predicted by the bending stiffness (i.e. $\sim\varphi^{-4/3}$). This short-range behavior is likely the result of both intra-aggregate branching which decreases correlations over a length proportional to the average distance between branches (which range from $~5-6$ according to analysis Fig. \ref{fig: branch density}) as well as fluctuations at the microscale of the square lattice that are not captured by a continuum bending description (see Appendix \ref{appendix: bending entropy}). The second is a longer-range regime ($s\gtrsim 10$) where the linear decay of tangent correlations suggests a persistence length that follows the predicted $\sim \varphi^{-4/3}$ scaling. This suggests that, notwithstanding branching along the aggregate backbone, the long range bending fluctuations are well described by an effective semi-flexible ribbon whose backbone orientation fluctuates over length scale $\sim \ell_{\rm p}$ that grows as frustration is reduced and domain width increases.

Evaluating the partition function for a worm-like chain of fixed backbone length $L_{\rm b}$ (see Appendix \ref{appendix: bending entropy}) it is straightforward to compute the entropic contribution from bending fluctuations:
\begin{equation}
\label{eq: sbend}
S_{\rm bend}\simeq\frac{\bar{N} k_{B}}{2}\bigg[1+\ln\frac{2\pi \lambda}{\ell_{\rm p}}\bigg]
\end{equation}
where $\bar{N}=L_{\rm b}/\lambda$ defines the effective number of bendable segments, and $\lambda \approx a$ is a microscopic cutoff for bending fluctuations.  Notably, this model predicts that the conformational entropy is extensive in the contour length of a self-limiting ribbon domain.  Assuming the ribbons to be roughly rectangular with constant thickness, the number of subunits per ribbon is simply $n = W_*L_{\rm b}/a^2$ and that the bending entropy of the bulk phase is neglible, we have a specific entropy difference due to conformational fluctuations
\begin{equation}
\Delta s_{\rm bend} \sim \frac{k_{\rm B} a}{W_*},
\end{equation}
where we are neglecting the much weaker logarithmic dependence encoded in the persistence length of domains.



\subsection{Translational entropy}

\begin{figure}[ht!]
\centering
\includegraphics[width=0.5\textwidth]{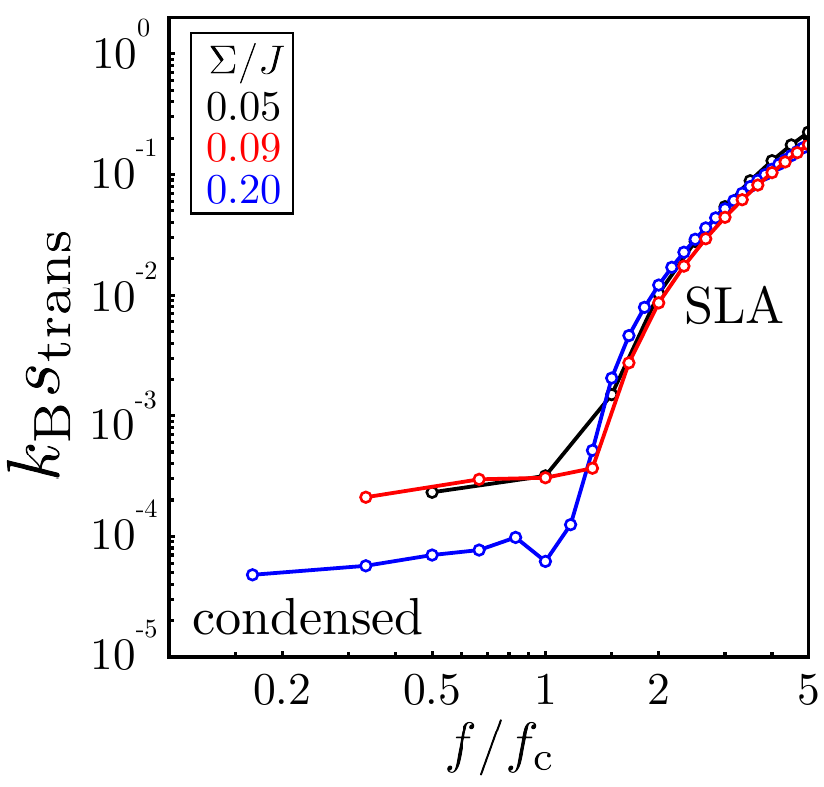}
\caption{Translational entropy per subunit measured over a range of frustration sweeping across the condensation transition. At weak frustration ($f/f_{\rm c}<1$) the translational entropy of the condensed phase is negligible and approximately constant. At strong frustration ($f/f_{\rm c}>1)$ the entropy of the self-limiting (SLA) phase increases rapidly with frustration. }\label{fig: translational entropy}
\end{figure}

In addition to the conformational entropy, any system of geometrically frustrated assembly will have translational entropy related to the discrete positioning of subunits. Notably, translational entropy in aggregating systems is size-dependent, in effect, because subunits within a single aggregate share a single center-of-mass degree of freedom. This translational entropy is encoded within the aggregate mass distribution, $\phi_n$, which denotes the total area fraction of the system contained within aggregates of size $n$. In the context of ideal aggregation theory~\cite{HaganGrason,israelachvili1976theory}, the mixing entropy of the distribution can be written as
\begin{equation}
S_{\rm trans.}=-\frac{k_{\rm B} A_{\rm tot}}{a^2} \sum_n \frac{\phi_n}{n}\Big(\ln \frac{\phi_n}{n}-1\Big),
\end{equation}
where $A_{\rm tot}$ is the total system area.
In particular, we are concerned with the translational entropy of {\it aggregated} structures, i.e. for $n > n_m$ where $n_m$ is the threshold value of cluster size delineating dispersed from aggregated structures. Throughout this work, we will take $n_m=9$, as this aligns with the approximate location of the minima of the aggregate concentration separating the monomer and SLA populations over the range of conditions considered here~\cite{HackneyPhysRevX.13.041010}. Summing over this part of the distribution and dividing by the number of aggregated subunits $\Phi_{\rm agg} A_{\rm tot}/a^2$, we have the {\it specific translational entropy of aggregated subunits}:
\begin{equation}
\label{eq: strans}
s_{\rm trans.}=-\frac{k_{\rm B}}{\Phi_{\rm agg}}\sum_{n>n_m} \frac{\phi_n}{n}\Big(\ln \frac{\phi_n}{n}-1\Big)
\end{equation}
which is straight forwardly computed from numerical simulations. In Figure \ref{fig: translational entropy}, we show the specific translational entropy of aggregated subunits, across the condensation transition, for several different values of $\Sigma/J$~\footnote{For this analysis, we use a constant value of aggregation threshold of $n_m = ??$}. From this, we see that, below the critical frustration when assembled subunits have condensed into a single bulk structure, the specific translational entropy of aggregates plummets to negligible values relative to the self-limiting state. This reflects the fact that, in the condensed state, the degrees of freedom of a single collective center of mass are distributed among a macroscopically large number of subunits in the structure, whereas in the self-limiting state, the translational entropy of aggregates is shared by a finite number of subunits, leading to a much larger specific entropy.


We estimate the specific translational entropy of self-limiting aggregates by assuming that the sum over the mass distribution in eq. (\ref{eq: strans}) is dominated by the peak in the distribution, which we have shown previously~\cite{HackneyPhysRevX.13.041010} occurs at mass $n_*$ roughly equal to the mass of a square aggregate of ideal self-limiting width, i.e. $n_* \approx (W_*/a)^2$.  From this, we estimate the value of the specific translational entropy of self-limiting aggregates as:
\begin{equation}
s_{\rm trans.} \approx -\frac{k_{\rm B}}{n_*} \ln \Big( \frac{\Phi_{\rm agg}}{n_*} \Big)
\end{equation}
where we used the approximation $\phi_{n_*} \approx \Phi_{\rm agg}$.  We note that this approximation undercounts the entropy associated with the breadth of the mass-distribution of self-limiting aggregates, in particular the contributions from the exponential tail for which $n \gg n_*$. However these large aggregates contribute a relatively small amount of specific entropy due to their large mass and so the effect of their omission is negligible.

Taking the translational entropy of the bulk state to be zero, this gives us the estimate for specific entropy change 
\begin{equation}
\Delta s_{\rm trans.} \propto 1/n_*
\end{equation}
which, neglecting logarithmic factors, also falls off with a power-law with self-limiting size, $W_*$, like the conformational entropy estimate.


\subsection{Entropic dependence of self-limiting to bulk transition}

The prior two sections give estimates of the greater specific entropy of self-limiting aggregates relative to the bulk aggregates.  We estimate that both conformational entropy (bending fluctuations) and translational entropy per subunit of self-limiting aggregate decrease with self-limiting width (albeit with different powers of $W_*$), in effect due to the larger number of subunits that share the same effective degrees of freedom in larger aggregates.  Here, we use these estimates to explore the effect that this excess entropy of the self-limiting aggregates has on the SLA to bulk transition and show that its main role is to depress the value of the critical frustration, $\varphi_c$, separating the self-limiting and defect bulk phases.  

\begin{figure}[ht!]
\centering
\includegraphics[width=0.5\textwidth]{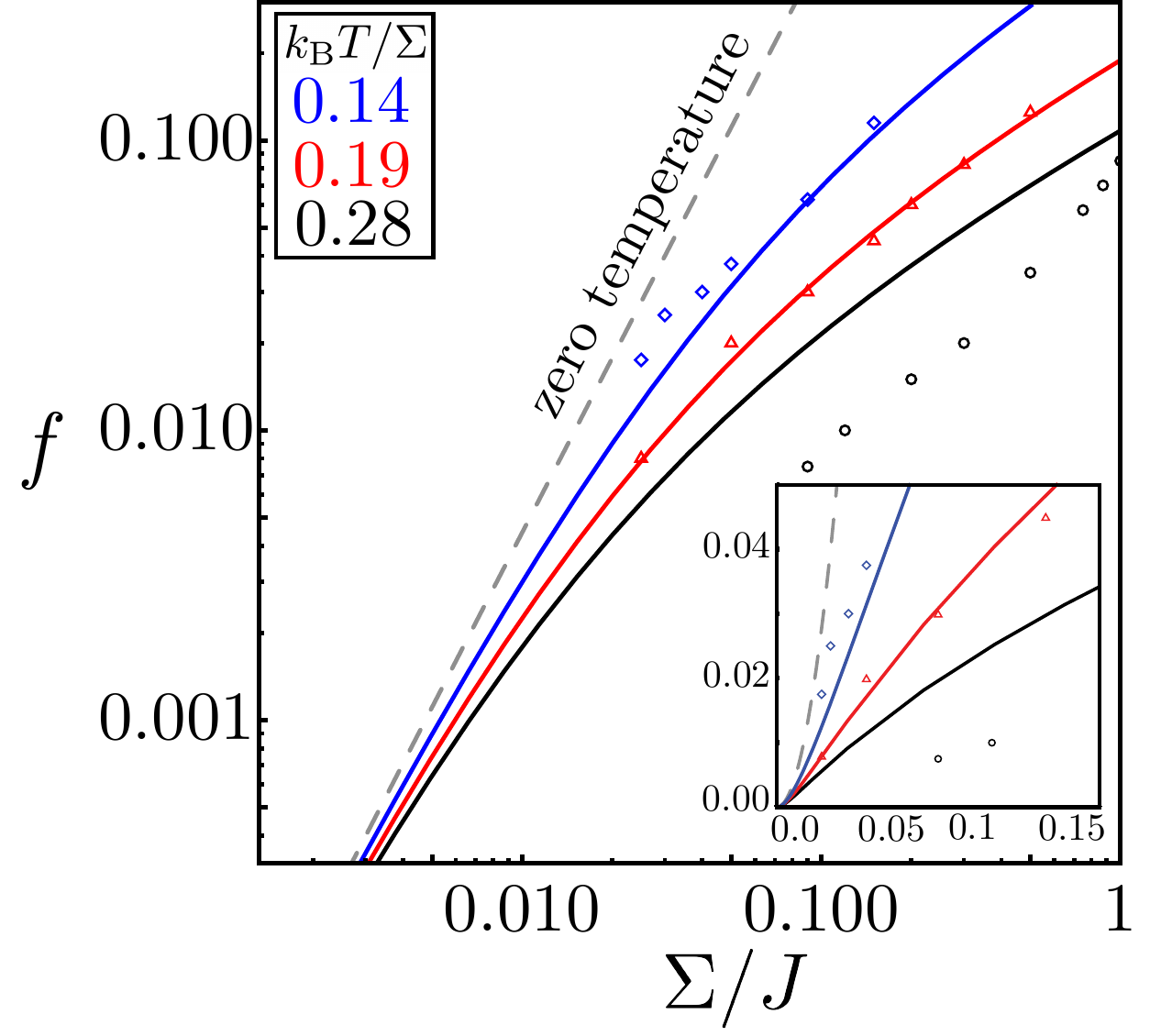}
\caption{Effect of entropy on the critical frustration. 
Dashed line shows the zero temperature prediction and the solid lines show theory prediction for the critical frustration as a function of cohesion-to-stiffness as temperature is varied. The open markers correspond to the numerically measured value of critical frustration. Inset shows same data on non-log plot to emphasize how far from quadratic scaling regime the numerical data falls.}\label{fig: critical frustration}
\end{figure}

For simplicity we consider only one of the two entropic contributions and its dependence on frustration and cohesion to stiffness ratio.  Based on an estimate of effective bond length $\lambda \approx 10 a$ from Fig.~\ref{fig: persistence length} in eq. (\ref{eq: sbend}) and values of specific entropy in Fig.~\ref{fig: translational entropy}, we expect translational entropy to dominate the specific entropy of self-limiting aggregates for the range of simulation parameters explored here.  Thus, using $n_* \approx (W_*/a)^2$ and $W_* \sim \varphi^{-2/3} (\sigma/J)^{1/3}$ we estimate a scaling form for the specific entropy change in the self-limiting to bulk transition
\begin{equation}\label{eq: delta entropy}
\Delta s\sim \frac{k_B}{n_*}=k_B a^2 \bigg(\frac{\varphi^2}{\sigma/J}\bigg)^{2/3},
\end{equation}
where we neglect the variation of the logarithmic term $\ln (\Phi_{\rm agg}/n_*)$ near the transition.  Substituting this into the difference in specific free energy between self-limiting and condensed states in eq. (\ref{eq: phase boundary}) we have:
\begin{equation}
\frac{\Delta F}{JA}=C_0 \bigg(\frac{\sigma}{J}\bigg)^{2/3}\varphi^{2/3}-C_1\varphi-C_{\rm T} \frac{k_B T}{\sigma}\bigg(\frac{\sigma}{J}\bigg)^{1/3}\varphi^{4/3},
\end{equation}
where $C_{\rm T}$ is a numerical prefactor.
It is straightforward to solve $\Delta F( \varphi)=0$ for $\varphi_c$, and the solutions are plotted in Figure \ref{fig: critical frustration}, where we convert to dimensionless frustration, $f_c$, to compared to simulation results for three values of reduced temperature $k_B T/\Sigma$.  Here a value of $C_T \simeq 45$ was chosen to best match the values extracted from simulation.  Notably the curves $f_c(\Sigma/J)$ indeed appear to show a roughly linear dependence on the cohesion to stiffness in the range of simulated values, and further shift to increasingly lower values as the reduced temperature $k_B T/\Sigma$ is increased. This is consistent with trends observed in the simulations.  Physically, the shift to low values $f_c(\Sigma/J)$ with increased temperature is a result of the relatively larger specific entropy of the self-limiting aggregates.  Indeed the large gap between the ground state $T=0$ value of $f_c$ (dashed line) and the values of $k_B T/\Sigma$ for the simulations (solid curves) imply that thermal effects are quite strong for simulated assemblies, and that the excess entropy of self-limiting aggregates over bulk states substantially accounts for the broad window of self-limiting assembly exhibited in our simulations.  

This mean-field type calculation further allows us to estimate the scale at which thermal effects become strong.  It is straightforward to show the perturbative correction to the $T=0$ critical frustration value at low-temperature
\begin{equation}
\varphi_{\rm c} ( T \ll J)\simeq \varphi_{\rm c} (T=0) \bigg[1 - \frac{4C_T C_0}{3C_1^2} \frac{ k_B T} {J}  \bigg] 
\end{equation}
where $\varphi_{\rm c} (T=0) = (C_0/C_1)^3 (\sigma/J)^2$ is the $T=0$ ground-state transition value.  We note that our simulations results are carried out for fixed values $k_B T/\Sigma$, and hence, increasing cohesion to stiffness ratio (i.e. $\Sigma/J$) corresponds to increasing values of $k_B T/ J$.  In the large temperature limit, we find:
\begin{equation}
\varphi_{\rm c} (k_B T \gg J) \sim \Big( \frac{k_B T}{\sigma} \Big)^{2/3} \sqrt{\sigma/J} .
\end{equation}
Hence, we expect a crossover from $\varphi_{\rm c} \propto (\sigma/J)^2$ to $\varphi_{\rm c} \propto (\sigma/J)^{1/2}$ around a cohesion scale inversely proportional to the fixed value of reduced temperature $k_B T/\Sigma$.  From Fig.~\ref{fig: critical frustration} and the simple mean field estimate, we expect that this crossover occurs in the range of $\Sigma/J \approx 10^{-2}-10^{-1}$.  Based on this, we conclude that the apparently linear dependence of critical frustration on $\Sigma/J$ may in fact be a manifestation of a broad crossover between an athermal regime to a strongly thermal regime from low to high values of cohesion to stiffness.  

We note by inspection of Fig.~\ref{fig: critical frustration}, that accessing the energetically-dominated athermal regime of the lattice GFA model, where $\varphi_{\rm c} \propto (\sigma/J)^2$, would likely require exceptionally low values of $\Sigma/J \lesssim 10^{-3}$ and correspondingly minuscule values of $f$. However, restrictions on the periodic boundary conditions (i.e. where $L$ is an integer multiple of $2/f$) require prohibitively large lattice sizes to access this range of ultra-low frustration.


\section{Conclusion and Discussion}\label{section: conclusion}

In this article, we studied a 2D lattice model of geometrically frustrated assembly to explore the interplay between translational degrees of freedom and frustration over the entire range of subunit concentration. This was done with a particular focus on understanding the various thermodynamic phase transitions separating the dispersed/self-limited aggregate phase from the topologically defective macroscopic bulk phase.

\begin{figure}[ht!]
\centering
\includegraphics[width=0.5\textwidth]{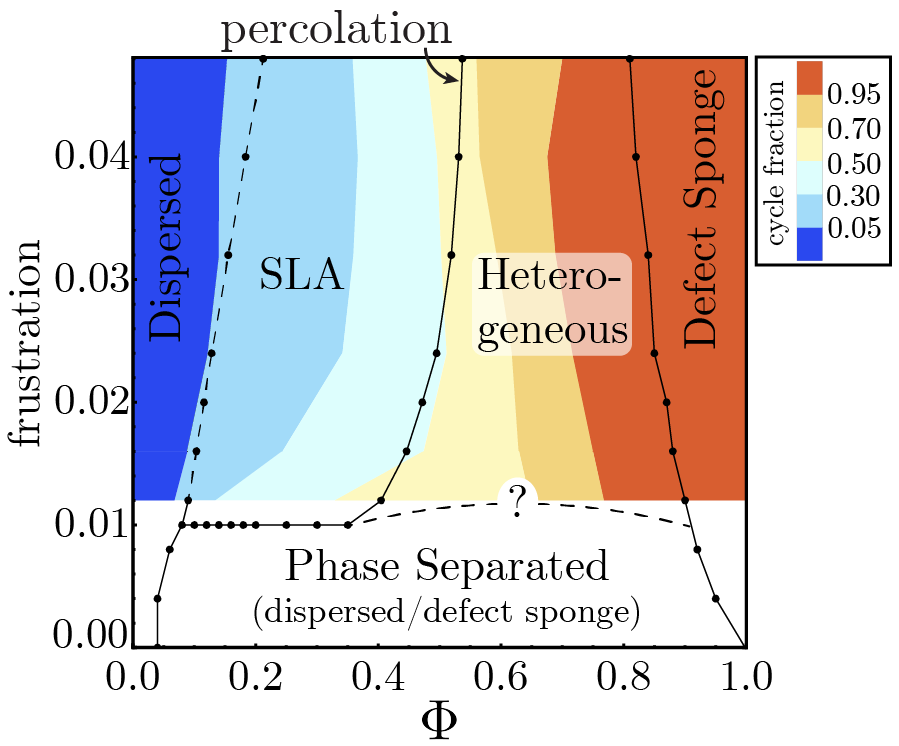}
\caption{Frustration-concentration plane of the phase space of geometrically frustrated assembly. Boundary between phases compiled from simulation results presented throughout this article. Solid dots correspond to individual simulations and the lines drawn between them included to guide the eye. Dashed line corresponds to the presumed continuation of the phase separation connecting low and high concentration. Simulations shown here run with $\Sigma/J=0.09$, $\beta J =40$ and $L=250,500$.}\label{fig: gfa phase}
\end{figure}

In section \ref{section: percolation}, we studied how, at strong frustration (i.e. above the phase separation binodal), the GFA model evolves from an increasing concentration of shape-fluctuating, finite-width domains above the aggregation transition to a heterogeneous percolated network at intermediate concentrations, and ultimately to a relatively uniform ``sponge'' of evenly spaced vortices, characterized by large-area cores.  We find that the location of this percolation transition shifts to somewhat higher concentration values with increased frustration, an observation that is consistent with both the decreasing overall mass of aggregates and an increase in their branch density with increasing frustration.   Additionally, near the percolation threshold, the structure of the sponge phase is quite heterogeneous and is predominantly characterized by large fluctuations in hole size and spacing, which eventually evolves at high concentration to evenly spaced holes of a size and separation that are selected by the energetics of frustration screening (i.e. a ``holey" Abrikosov lattice).  Notably, in characterizing the loopy versus tree-like structure of the self-limiting domain, we find a very broad crossover, extending from concentrations well below the percolation point up to optimal sponge density.  This analysis is summarized in Fig. \ref{fig: gfa phase}, showing the ``short cycle" fraction of aggregated clusters as a function of concentration.  This rather gradual crossover between extended finite-width, worm-like domains and defective loops of well-defined size suggests a more complex scenario of coexistence between these morphologies at strong frustration in which defect loops effectively condense in small clusters onto extended tree-like ribbon domains (akin to grapes on the vine).  Collectively, these ``two-phase" aggregates ultimately mix into a single network, reflecting relatively even proportions of tree-like and ``short loop" structures.  Increasing concentration generically favors the denser defect-loop morphology, ultimately leading to quasi-uniform sponge.

In section \ref{section: condensation}, we studied contributions to the assembly entropy and how these tend to thermodynamically stabilize self-limiting aggregates relative to the bulk condensed state.  We considered simple models of conformational and translational entropy of self-limiting aggregates, which predict that the specific entropy of both effects depends on a reciprocal power of the self-limiting domain size. Based on the scaling of domain size on frustration and cohesion to stiffness ratio in the translational entropy, a mean-field model of the free energy difference between self-limiting and bulk states was used to predict the finite-temperature dependence on the critical frustration, $\varphi_c$, separating these states.  Comparison between this simple model and simulation results suggest that our GFA simulations are in a strongly thermal regime, in which favorable entropy of dispersed self-limiting aggregates substantially depresses $\varphi_c$ below its expected $T=0$ scaling $\varphi_c(T=0) \sim (\sigma/J)^2$.  Consistent with previous observations~\cite{HackneyPhysRevX.13.041010}, this model shows that $\varphi_c$ decreases with increasing temperature, in effect a thermal melting of condensed, spongey bulk states into a disperse suspension of extended and finite-width aggregates. The apparent linear dependence on cohesion to stiffness observed for $\varphi_c$ at fixed temperature is consistent with a broad crossover regime from square dependence to a square-root dependence at low versus high values of $\sigma/J$, although direct numerical evidence of the $T=0$ scaling appears likely out of the accessible range of current simulations. 

In summary, this sheds light on key aspects of geometrically frustrated assembly at finite temperature and concentration. Firstly, this shows that, while energetic considerations of elastic cost of frustration and cohesion control the size and structure of the competing states of aggregation, conformational and configurational entropy play a significant, and heretofore underappreciated, role in stabilizing states of self-limiting aggregation. Additionally, this study addresses the fate of self-limiting assembly far above the conditions of ideal aggregation, at concentrations where aggregates are strongly interacting.  We demonstrate the concentration bound on the range of conditions that support self-limited assembly, and study how the system evolves from a dispersed solution of aggregates to a relatively ordered and uniform bulk sponge via an intermediate percolating and heterogeneous network mesophase.  This scenerio provides a framework for connecting prior studies of bulk and dilute geometrically frustrated assemblies that were previously treated as unrelated. For example, much work has been done studying the bulk defect phase of bend-nematic liquid crystals~\cite{fernandez2021hierarchical,kotni2022,subert2024achiral}. Meanwhile, the same mechanism of frustration has been theorized to induce self-limitation under dilute conditions~\cite{HackneyPhysRevX.13.041010,tanjeem2022focusing,sullivan2024self}. Thus, the framework outlined here could provide a useful guideline for extending earlier experimental work to the dilute phase, thereby opening up a new avenue for designing novel functional material. Conversely, the pathway could go the other way, helping to extend work done on dilute self-assembly to build complex bulk material from the ground up.  Additionally, we note several previous studies of similar phase behavior in the context of colloidal-gels of attractive particles~\cite{Zaccarelli_2007}.  While many models of these systems assume that the locally aggregated structures form far from equilibrium (i.e. via rapidly quenching~\cite{Zhang_2019}), there has been some investigation into the possibility that at least strongly-metastable favored local structures form as a result of geometric frustration of polytetrahedral packing of a small number of sticky spheres~\cite{Royall_2008, Royall_2021}.  In this context, we anticipate that the lattice GFA model may indeed by a useful analog of such systems in which the frustration may be systematically tuned to adjust the equilibrium dimensions of favored local structures (e.g. domain sizes and inter-defect space). Thus allowing one to probe the statistics, structure and mechanics of the emergent networks.

While the work presented above provides a complete picture of the effect of frustration at finite temperature on assembly over the full range of concentration, several aspects remain to be understood. For one, dispersed monomers and self-limiting aggregates are two types of low concentration solution phases (separated only by a pseudo-critical aggregation crossover) it is far from clear if or how the percolation intersects or merges with the phase separation binodal. Here, there could be a tri-critical point similar to that studied in the Blume-Emory-Griffiths model of phase separation in thin films of super-fluid He$^3$-He$^4$ mixtures~\cite{blume1971ising,berker1979superfluidity}. Additionally, the observation that the percolating clusters consist of small, dense regions connected by thin ribbons suggests the possibility of a secondary binodal where the heterogeneous sponge phase is viewed as a coexistence between the homogeneous sponge and self-limiting assembly.  Our current speculation (indicated graphically in Fig. \ref{fig: gfa phase}) is that such a critical point could be hidden within the window between the percolation threshold and the optimal density of the uniform vortex array. However, equilibration of the nearly bulk, yet strongly-fluctuating, configuration of states in this regime is especially challenging.

Beyond the transition between strong- and weak-frustration, the nature of the high frustration (i.e. $f\rightarrow 1/2$) limit is particularly unclear. As frustration increases, the aggregate widths decrease until $W\sim a$. For the parameters used in Figure \ref{fig: gfa phase}, this happens around $\varphi\simeq0.06$. Above this, the aggregates are expected to form living polymeric chains of single subunit width. As geometric frustration only acts on closed cycles of connected bonds, these 1D spin chains could avoid the adverse effects of frustration by forming fractal network structures that are self-avoiding and tree-like. Although  the concept of a self-limiting aggregate becomes poorly defined in this limit and it is unclear what bearing this regime of phase space has on the behavior of physical assemblies, we speculate that certain conformational statistics, if not also critical phenomena, would likely be described within the universality class of branched lattice animals~\cite{lubensky1979statistics}.

\begin{acknowledgments}

The authors are grateful to Jon Machta and Christopher Amey for helpful and detailed discussions about MC simulation approaches and to Kyle Sullivan and Christian Koertje for valuable feedback on this manuscript. This work was supported by the National Science Foundation through Award No. NSF-DMR 2028885 and 2349818. All simulations were performed using the UMass Unity cluster located at the Massachusetts Green High Performance Computing Center.

\end{acknowledgments}

\appendix

\section{Local Orientational Ordering of Defect Holes}\label{appendix: orientational order}

\begin{figure}[ht!]
\centering
\includegraphics[width=0.5\textwidth]{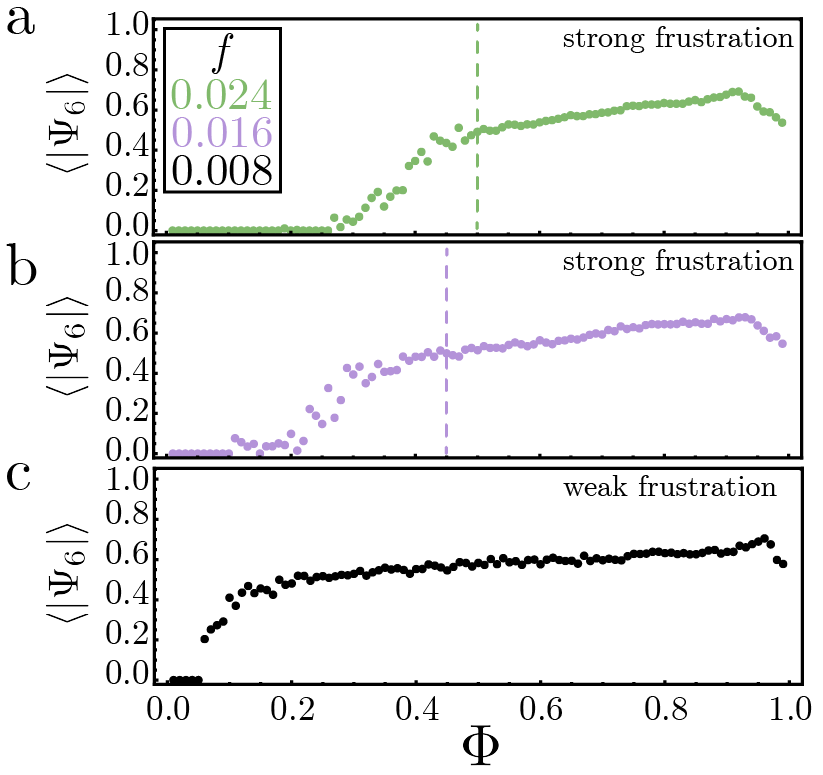}
\caption{Local orientational order at strong and weak frustration. Vertical dashed line represents location of percolation transition occurring at strong frustration. All simulations run with $\Sigma/J=0.09$, $\beta J =40$ and $L=250$.} \label{fig: local orientation}
\end{figure}

The orientational ordering of the intra-assembly holes can be quantified via the local hexatic order parameter\cite{prestipino2011hexatic}:
\begin{equation}
\Psi_{6}=\frac{1}{N_{nn}}\sum_{nn}e^{6i\theta_{nn}}
\end{equation}
which measures how closely the local arrangement of holes aligns with a perfect triangular lattice. Here $\theta_{nn}$ defines the angle that the vector between the center of two nearest neighboring holes makes with respect to the $\hat{x}$ direction and $N_{nn}$ is the number of nearest neighbors around a given hole. Averaged over every hole in an aggregate, the ensemble average $\langle \vert\Psi_{6}\vert\rangle$ assumes a value ranging from 0, corresponding to the total absence of orientational order, to 1, which corresponds to a perfect triangular lattice. This quantity is evaluated over the entire range of concentration for two different values of strong frustration (see Figure \ref{fig: local orientation}ab). Here, we observe an initial rise in local orientational order that coincides with the percolation transition. 
At weak frustration, we observe a similar increase in local orientational order, only here it coincides with the onset of phase separation, which happens at a lower value of concentration than the percolation transition (see Figure \ref{fig: local orientation}c). In both cases, the orientational order continues to increase with concentration from $\langle\vert\Psi_6\vert\rangle\sim 0.4$ to $\sim0.7$. While this suggests some amount of local orientational ordering, the overall effect is not very strong and there is no discernible differentiation between the heterogeneous, homogeneous and condensed states of assembly under these conditions. This suggests that any amount of apparent ordering is the result of the uniformity in hole size and spacing, rather than any strong hexagonal arrangement.

\begin{figure*}[ht!]%
\includegraphics[width=\textwidth]{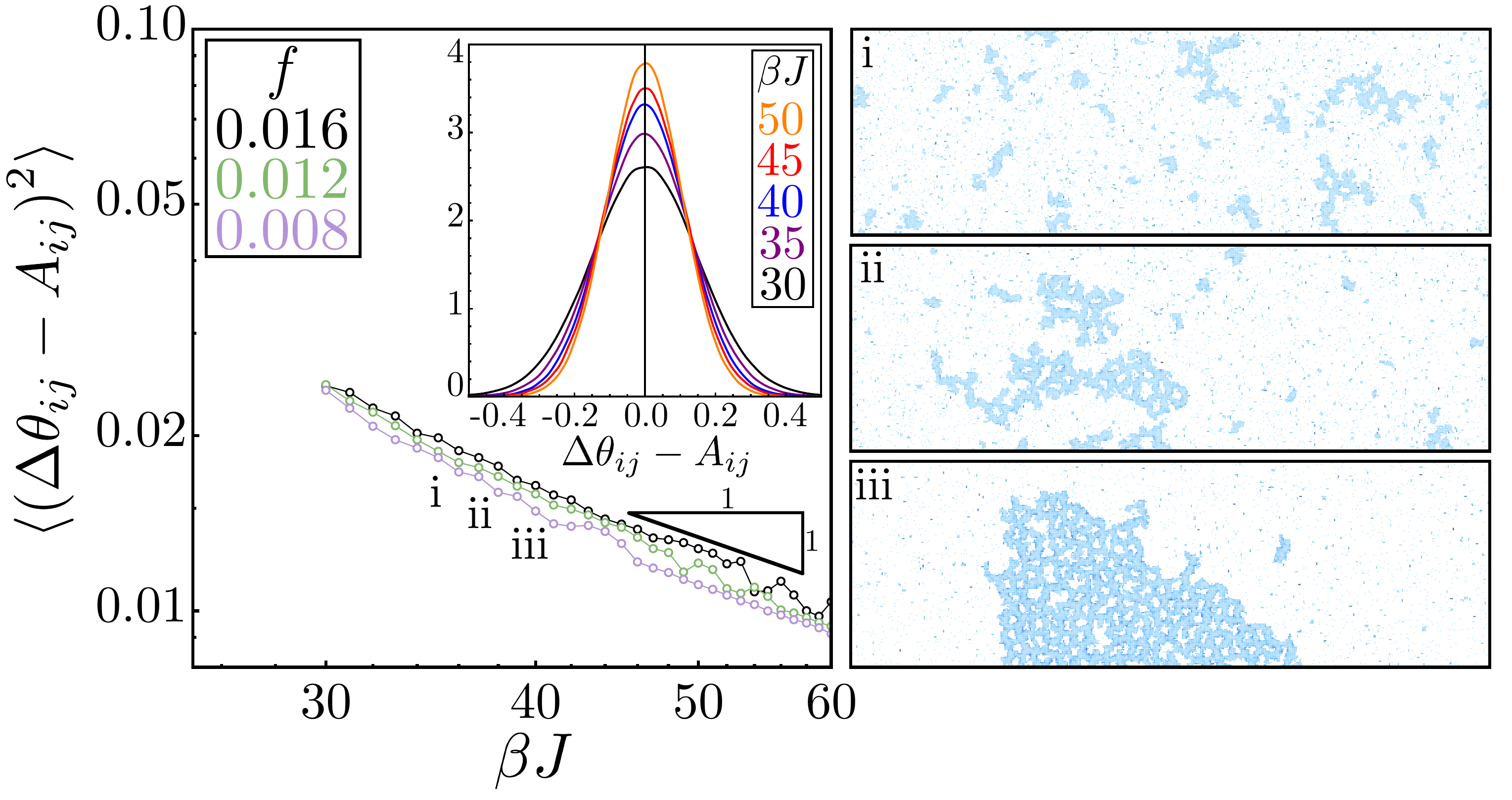}
\caption{Temperature dependence of the gauge invariant nearest neighbor spin fluctuation across phase transition. Simulations were run for three different values of frustration ($f=0.008$, $0.012$, and $0.016$) with $\Sigma/J=0.09$ and $\Phi=0.15$. The inset shows histograms of nearest neighbor phase difference as a function of temperature. This data was taken from the $f=0.008$ for several different temperatures crossing the transition. Example simulation snapshots (i-iii) are included for temperatures on either side of the transition with (i) illustrating the self-limiting state, (ii) illustrating simulation near the critical temperature and (iii) representing a phase separated bulk. Simulations shown here run with $\Sigma/J=0.09$, $L=250$. } \label{fig: phase fluctuations}
\end{figure*}
\section{Orientational Entropy}\label{appendix: orientational entropy}

The conformational entropy due to fluctuations in subunit orientation can be calculated within the Einstein crystal approximation~\cite{kittel2018introduction}, which assumes that the lattice link between each pair of neighboring subunits is an independent variable. The validity of this assumption can be immediately seen by looking at the distribution of the gauge invariant phase difference, $\Delta\theta_{ij}-A_{ij}$, along occupied links (see inset of Figure \ref{fig: phase fluctuations}) and observing that they are Gaussian in nature. Furthermore, this suggests that we can obtain the entropy of these fluctuations by taking the log of their variance, i.e.
\begin{equation}
S\simeq\ln\langle(\Delta\theta-A)^2\rangle
\end{equation}
Noting that the Hamiltonian for this system (eq. (\ref{eq: lattice hamiltonian})) is quadratic in $\Delta\theta-A$ allows us to calculate the variance using the equipartition theorem~\cite{pathria1996statistical}. Therefore,
\begin{equation}
S\simeq-\ln\beta J
\end{equation}
which only depends on the ratio of temperature and spin-stiffness. Comparing this to the orientational entropy measured from a range of simulations sweeping across the transition (see Figure \ref{fig: phase fluctuations}), we find good agreement in both the self-limiting and condensed phase. Thus, we see that fluctuations in subunit orientation are largely independent of phase and frustration, suggesting that the change in orientational entropy across the transition should be negligible.

\begin{figure*}[ht]%
\includegraphics[width=\textwidth]{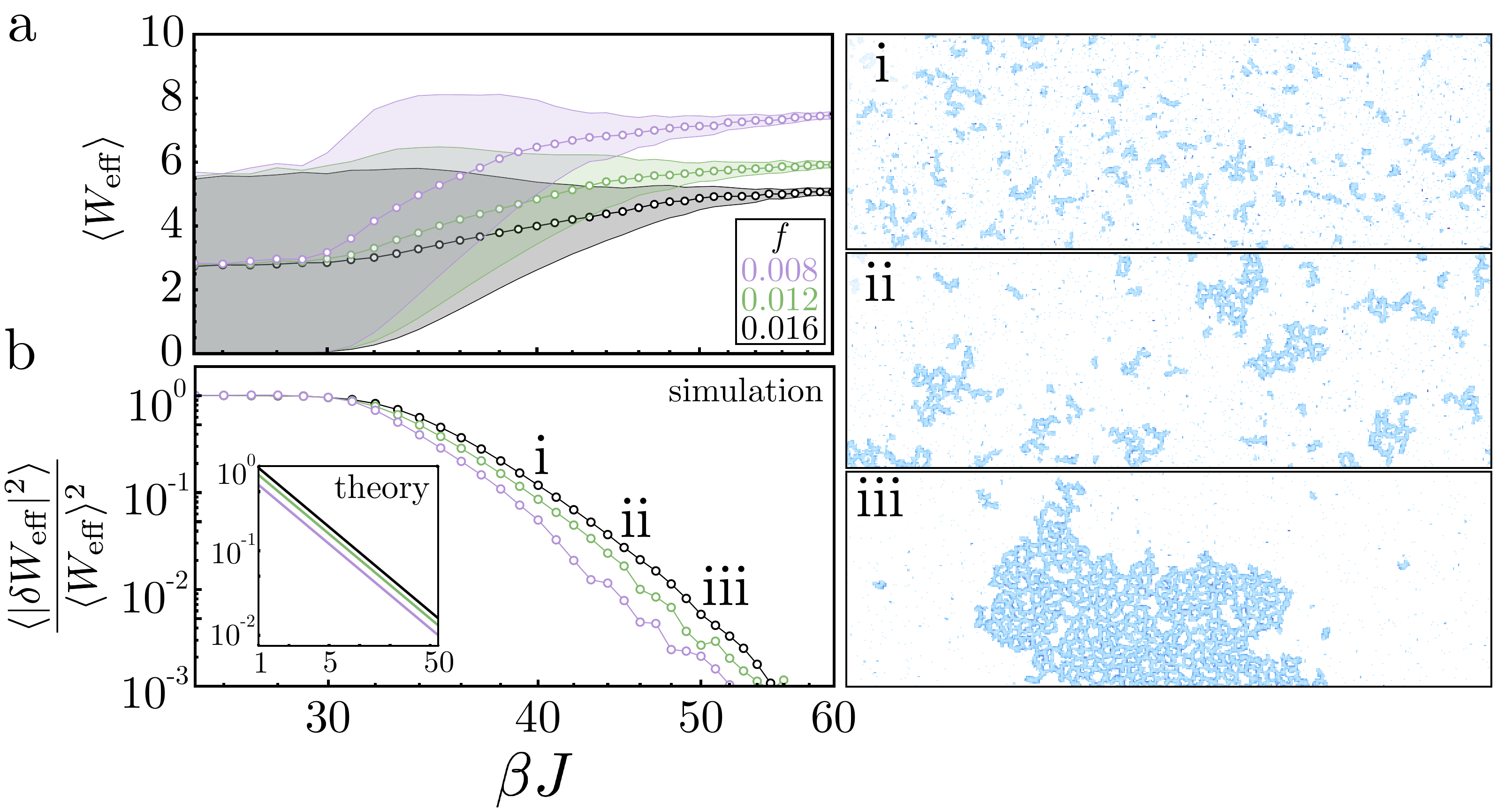}
\caption{Capillary fluctuations across phase transition. (\textbf{a}) Dependence of mean aggregate width on temperature. Simulations were run for three different values of frustration ($f=0.008$, $0.012$, and $0.016$) with $a\Sigma/J=0.09$ and $\Phi=0.15$. The bands of faded color denote the standard deviation of measured width distribution. (\textbf{b}) Variance of aggregate width distribution normalized by mean aggregate width. The inset shows the harmonic correction to the mean field theory (eq. (\ref{eq: harmonic correction})).  Example simulation snapshots (i-iii) are included for temperatures on either side of the transition with (i) illustrating the self-limiting state, (ii) illustrating simulation near the critical temperature and (iii) representing a phase separated bulk. } \label{fig: capillary fluctuations}
\end{figure*}

\section{Capillary Fluctuations}\label{appendix: capillary fluctuations}
 
 The aggregates observed in our finite temperature simulations exhibit noticeable capillary fluctuations in aggregate width. The importance of these fluctuations can be evaluated by considering an infinitely long ribbon of varying width, described by $W(z)=W_*+\delta W$, where $z$ is the position along the aggregate backbone and $\delta W$ is the deviation of the width away from $W_*$. The variation in aggregate width leads to a change in free energy~\cite{spivack2022stress}:
\begin{equation}\label{eq: delta F}
\Delta F[\delta W(z)]\simeq \frac{1}{2}\int\rm dz \bigg[\frac{\Sigma}{2}(\frac{\partial}{\partial z}\delta W)^2+M(\delta W)^2\bigg]
\end{equation}
where
\begin{equation}
M(W_*)=\frac{\partial^2}{\partial w^2}\bigg[W\epsilon(W)\bigg]\vert_{W_*}=J\pi^2 \varphi^2 W_{*}.
\end{equation}
Introducing the Fourier transform width fluctuation:
\begin{equation}
\delta W(k)=\int \rm dz e^{ikz}\delta W(z)
\end{equation}
we can re-write eq. (\ref{eq: delta F}) as:
\begin{equation}
\Delta F[\delta W(z)]=\frac{1}{2}\int\frac{\rm dk}{2\pi}\bigg[\frac{\sigma k^2}{2}+M\bigg]\vert\delta W(k)\vert^2
\end{equation}
The equipartition theorem then tells us that:
\begin{equation}
\langle\vert\delta W(k)\vert^2\rangle=\frac{\beta^{-1}}{\frac{\sigma k^2}{2}+M}
\end{equation}
Fourier transforming back to real space, we get the thermal width fluctuation~\cite{spivack2022stress}:
\begin{equation}
\langle\vert\delta W(z)\vert^2\rangle=\frac{\beta^{-1}}{\sqrt{2\sigma M}}=\frac{\beta^{-1}}{\sqrt{2\sigma J \pi^2 \varphi^2 W_{*}}}
\end{equation}
Comparing this to the square of the mean width, $\langle W\rangle^2$, we expect that capillary fluctuations will be negligible when:
\begin{equation}\label{eq: harmonic correction}
\frac{\langle\vert\delta W\vert^2\rangle}{\langle W_* \rangle^2}=\frac{(\beta J)^{-1}}{\sqrt{2\sigma/J}}\frac{W_*^{-5/2}}{\pi \varphi}\ll 1
\end{equation}

The theory discussed above describes harmonic corrections to the mean field theory, which assumes that $\langle W_* \rangle$ is independent of temperature and only depends on the dimensionless parameters $f$ and $\Sigma/J$. While this is true in the limit of large $\beta J$, the mean width that we measure via simulation varies with temperature (see Fig \ref{fig: capillary fluctuations}a). This dependence is largely related to the tendency of increased temperature to drive aggregates towards dissolution, thereby lowering the fraction of subunit mass contained within an aggregate~\cite{HackneyPhysRevX.13.041010}. As this effect is not captured by the mean-field theory, it is difficult to make a direct comparison between theory and simulation. However, the interpretation of the quantity $\langle\vert\delta W\vert^2\rangle/\langle W_*\rangle$ is still the same and our numerical results are at least in qualitative agreement with the theory (see Fig. \ref{fig: capillary fluctuations}b). Perhaps more presciently, the simulation data shows that\textemdash for temperatures near the SLA to bulk transition\textemdash the magnitude of width variation is small compared to the mean. This suggests that capillary fluctuations do not provide the dominant correction to the mean field prediction for the critical value of frustration.
\section{Bending Energy}\label{appendix: bending energy}
\begin{figure}[ht!]
\centering
\includegraphics[width=0.3\textwidth]{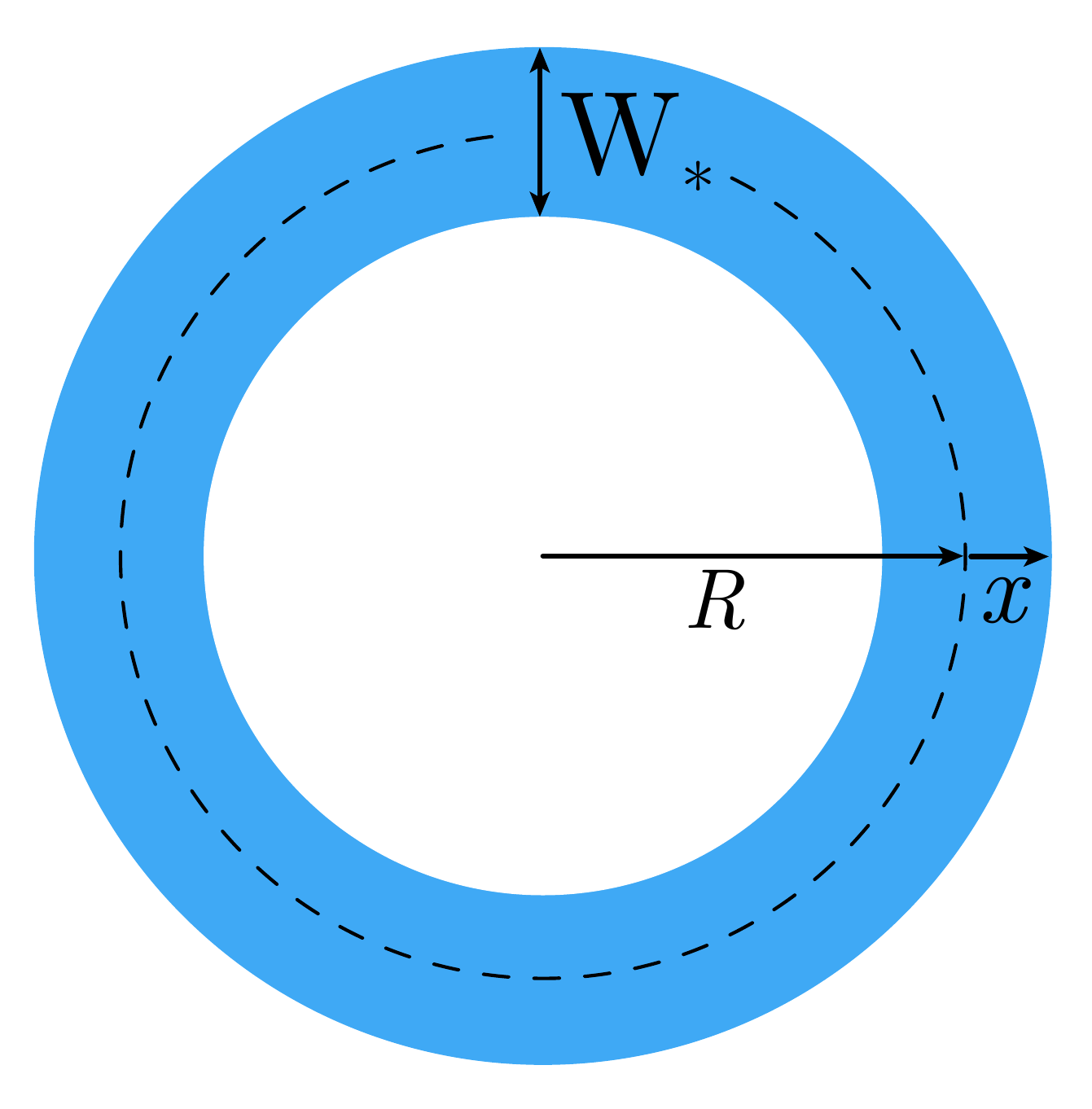}
\caption{Schematic illustration of an annular domain with midpoint radius, $R$, and finite width, $W_*$, that we use to calculate the bending energy as a function of curvature.}\label{fig: annulus}
\end{figure}

In order to find the bending modulus of our self-limiting aggregates, we calculate the curvature dependent term of the excess energy density of an annular domain with finite width $W_*$ (see Fig. \ref{fig: annulus}). To start, we introduce the cross-sectional coordinate,
\begin{equation}
r=R+x
\end{equation}
of our annular domains in terms of the midpoint radius, $R$. This allows us to write down the solution to the Euler-Lagrange equation (eq. (\ref{eq: euler-lagrange})) for an annular domain as:
\begin{equation}
\theta=s\tan^{-1}\bigg(\frac{y}{R+x}\bigg)
\end{equation}
where $s$ defines the winding around the annular domain. For convenience, we define the gauge field over the annulus as:
\begin{equation}
\mathbf{A}=\pi \varphi(R+x)\hat{\phi}
\end{equation}
Using these, we can calculate the energy via:
\begin{equation}
\begin{split}
H&=\frac{J}{2}\int_0^{2\pi}\int_{-W_*/2}^{W_*/2}\vert \nabla\theta-\mathbf{A}\vert^2\\
&=J\pi ARW_*+\frac{J\pi}{12 R}(B+C)W_*^3\\
\end{split}
\end{equation}
where, for convenience, we have defined:
\begin{equation}\label{ABC}
\begin{split}
A&=\frac{s^2}{R^2}-2\pi \varphi s+\pi^2 \varphi^2 R^2 \\
B&=2\pi^2 \varphi^2 R^2-\frac{2 s^2}{R^2} \\
C&=\frac{3s^2}{R^2}+\pi^2 \varphi^2 R^2 \\
\end{split}
\end{equation}
Dividing by the area and minimizing with respect to $s$, we find the optimal phase winding:
\begin{equation}
s=\frac{\pi \varphi R^2}{1+\frac{W_*^2}{12 R^2}}\simeq \pi \varphi R^2\bigg(1-\frac{W_*^2}{12 R^2}+\mathcal{O}(x^4)\bigg)
\end{equation}
Substituting these back into the Hamiltonian and introducing the parameterizing the infinitesimal arc length as $\rm d\theta=\frac{\rm d s}{R}$, we find the bending energy as a function of curvature:
\begin{equation}
\begin{split}
H&=\frac{J}{2}\bigg(\frac{B+C}{R^2}\bigg)\int_{-W_*/2}^{W_*/2}\int x^2 \rm dx~\rm ds \\
&=\frac{J\pi^2\varphi^2W_*^5}{72}\int \frac{\rm ds}{R^2}
\end{split}
\end{equation}
This allows us to read off the bending modulus
\begin{equation}
B=\frac{J\pi^2\varphi^2W_*^5}{72}
\end{equation}
given in equation (\ref{eq: bending modulus}).
\section{Bending Entropy}\label{appendix: bending entropy}

In order to calculate the conformational entropy associated with bending, we treat the self-limiting aggregates as 1D chains of $\bar{N}=L_{\rm b}/\lambda$ segments. Here, $L_{\rm b}$ is the total length of the aggregate backbone and $\lambda$ is a microscopic cutoff related to the shortest bendable unit. Thus, we can rewrite the continuum bending energy defined in equation (\ref{eq: bending energy}) as the discrete sum:
\begin{equation}
H_{\rm bend}=\frac{\beta B}{2}\sum_{i=0}^{\bar{N}}\frac{(\theta_{i+1}-\theta_{i})^2}{\lambda}
\end{equation}
where $\theta_i$ is the angle each segment makes with respect to the $x$-axis. Using this, we can write the partition function as
\begin{equation}
Z=\prod_{i=1}^{\bar{N}}\int d\theta_{i}e^{-\frac{\beta B}{2}\sum\frac{(\theta_{i+1}-\theta_{i})^2}{\lambda}}
\end{equation}
Assuming that the orientation of each segment is independent, this can be expressed as a product of $\bar{N}$ Gaussian integrals and evaluated to find:
\begin{equation}
Z=\bigg(\frac{2\pi \lambda}{\ell_{\rm p}}\bigg)^{\frac{\bar{N}}{2}}
\end{equation}
where $\ell_{\rm p}=\beta B$ is the persistence length of the chain. From here, it is easy to obtain the free energy:
\begin{equation}
\begin{split}
F&=-k_{\rm B}T\ln Z \\
&=-\frac{\bar{N}k_{\rm B}T}{2}\ln \frac{2\pi\lambda}{\ell_{\rm p}}
\end{split}
\end{equation}
This can be differentiated to find the conformational entropy of bending given in equation (\ref{eq: sbend}):
\begin{equation}
S_{\rm bend}=-\frac{\partial F}{\partial T}=\frac{\bar{N}k_{\rm B}}{2}\bigg[1+\ln\frac{2\pi\lambda}{\ell_{\rm p}}\bigg]
\end{equation}
\begin{figure*}[ht!]
\includegraphics[width=\textwidth]{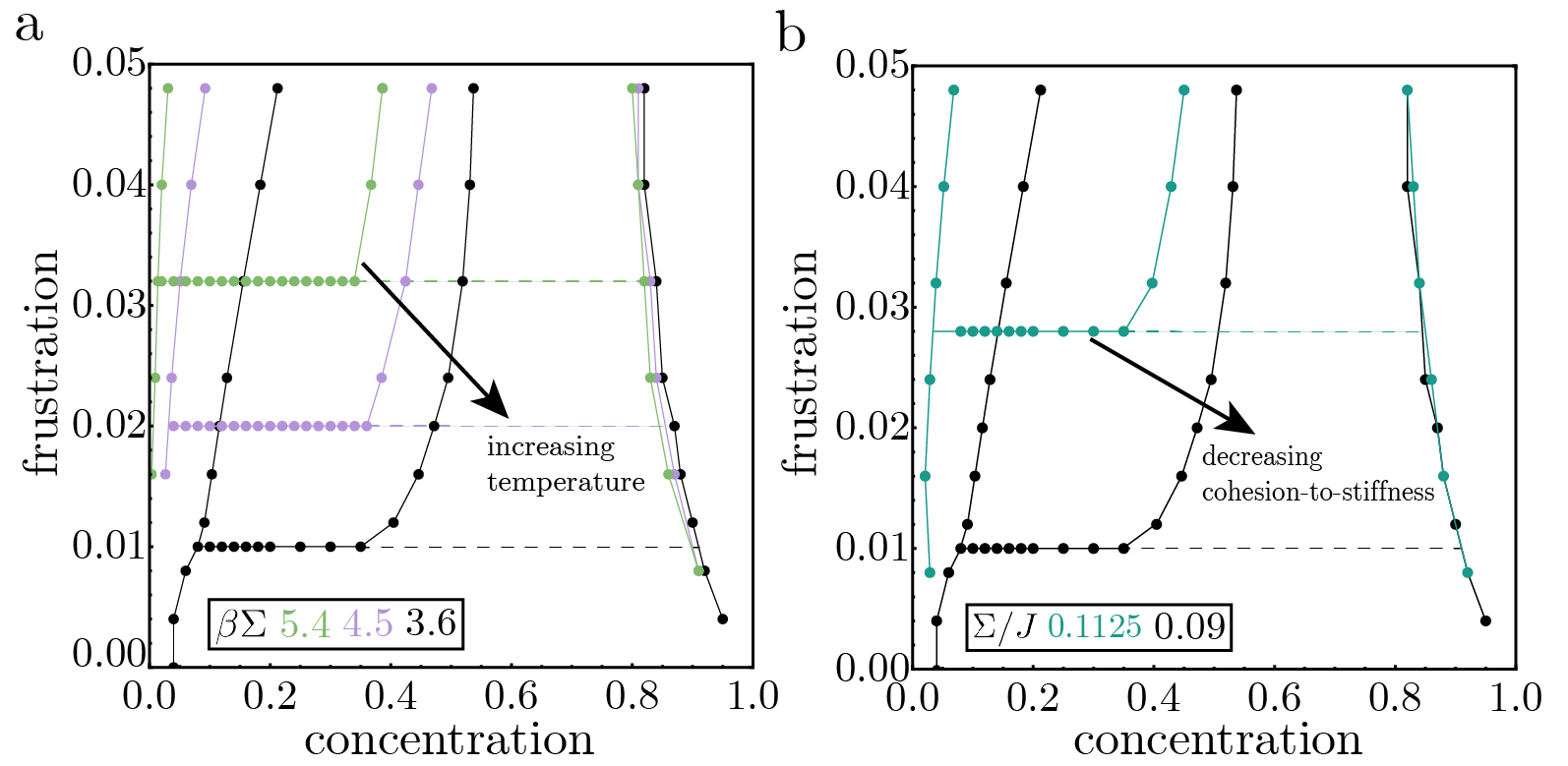}
\caption{Dependence of phase behavior on the temperature and cohesion-to-stiffness ratio. (\textbf{a}) The phase diagram of geometrically frustrated assembly is shown for several different values of the dimensionless temperature $\beta\Sigma$. Simulations run with fixed $\Sigma/J=0.09$ and $L=250,500,800$ depending on frustration. (\textbf{b}) The phase diagram of geometrically frustrated assembly is shown for several different values of the dimensionless temperature $\beta\Sigma$. Simulations run with fixed $\beta J=40.0$ and $L=250$. } \label{fig: phase diagram variations}
\end{figure*}
\FloatBarrier
\bigskip
\bibliography{apssamp}

\end{document}